\documentclass[twocolumn, twocolappendix]{aastex631}

\usepackage{booktabs}

\usepackage[shortlabels]{enumitem}
\usepackage{hyperref}
\usepackage[utf8]{inputenc}
\usepackage{svg}
\usepackage{booktabs}
\usepackage{makecell}  
\begin{document}

\title{Stellar Mass Census of $S^4G$ Local Bars}

\author[0009-0007-3523-5140]{Yasmin Cavalcante-Coelho}
\affiliation{Valongo Observatory, Federal University of Rio de Janeiro, 
Ladeira de Pedro Antônio 43, Rio de Janeiro, RJ 20080090, Brazil}
\affiliation{Kyungpook National University
80, Daehak-ro, Buk-gu, Daegu, Republic of Korea}
\email{ycoelho@ov.ufrj.br}

\author[0000-0003-3153-5123]{Karín Menéndez-Delmestre}
\affiliation{Centro Brasiliero de Pesquisas Físicas (CBPF), Rua Dr. Xavier Sigaud, 150 - Botafogo, Rio de Janeiro - RJ, 22290-180}
\affiliation{Valongo Observatory, Federal University of Rio de Janeiro, 
Ladeira de Pedro Antônio 43, Rio de Janeiro, RJ 20080090, Brazil}

\author[0000-0002-5857-5136]{Taehyun Kim}
\affiliation{Department of Astronomy, Yonsei University, 50 Yonsei-ro, Seodaemun-gu, Seoul 03722, Republic of Korea}
\affiliation{Kyungpook National University
80, Daehak-ro, Buk-gu, Daegu, Republic of Korea}

\author[0000-0002-0620-136X]{Arianna Cortesi}
\affiliation{Instituto de Física, Universidade Federal do Rio de Janeiro, Rio de Janeiro, RJ 21941-972, Brazil}

\author[0009-0005-1424-6604]{Natan de Isídio}
\affiliation{European Southern Observatory, Karl-Schwarzschild-straße 2, 85748 Garching bei München, Germany\\}

\author[0000-0003-2374-366X]{Thiago S. Gonçalves}
\affiliation{Valongo Observatory, Federal University of Rio de Janeiro, 
Ladeira de Pedro Antônio 43, Rio de Janeiro, RJ 20080090, Brazil}

\author[0000-0002-5496-4118]{Kartik Sheth}
\affiliation{Laboratoire d’Astrophysique de Marseille (LAM), UMR7326, CNRS/Aix Marseille Université, Technopôle de Marseille-Etoile, 38 rue Frédéric Joliot Curie, F-13388
Marseille Cédex 13, France}

\author[0000-0002-2808-1223]{Osbaldo Sánchez-García}
\affiliation{Departamento de Astronomia, Instituto de Astronomia, Geofísica e Ciências Atmosféricas da Universidade de São Paulo, Cidade Universitária, 05508-900 São Paulo, SP, Brazil}

\author[0000-0003-4675-3246]{Marco Grossi}
\affiliation{Valongo Observatory, Federal University of Rio de Janeiro, 
Ladeira de Pedro Antônio 43, Rio de Janeiro, RJ 20080090, Brazil}

\author[0009-0006-2067-3263]{Thiago Bueno-Dalpiaz}
\affiliation{Centro Brasiliero de Pesquisas Físicas (CBPF), Rua Dr. Xavier Sigaud, 150 - Botafogo, Rio de Janeiro - RJ, 22290-180}

\begin{abstract}

Bars are stellar structures commonly found in disk galaxies and important agents of secular evolution. Quantifying the stellar mass content of bars provides a physically motivated approach to assess how bar growth depends on galaxy mass and internal structure.
We perform a GALFITM multi-band mid-infrared two-dimensional decomposition of 344 barred galaxies from the Spitzer Survey of Stellar Structure in Galaxies ($S^4G$) in the 3.6 and 4.5 $\mu$m bands. 
We estimate the stellar mass associated with each structural component (i.e., disk, bulge, bar, and nuclear point source) and investigate how bar properties correlate with global galaxy characteristics. Our final models consist of 2-4 components per galaxy, with the simplest configuration (disk + bar) typically associated with low-mass systems.
We find that more massive galaxies with prominent bulges host more massive bars and higher median bar-to-total stellar mass ratios. These bulges generally present Sérsic indices below 2, consistent with a photometric signature of nuclear stellar disks.
Additionally, when we consider the relative bar lengths at a fixed galaxy mass, we find a  bimodal distribution where galaxies with bulges host relatively
longer bars compared to their bulgeless counterparts.
These findings could be the result of different evolutionary paths, possibly linked to different disk formation epochs, allowing bars more time to grow in size and mass. 
 Our study indicates that bar evolution may follow distinct pathways across different galaxy mass regimes, and is likely influenced by secular processes and environmental effects. 

\end{abstract}

\keywords{galaxies: nearby galaxies -- galaxies: barred galaxies -- galaxies: stellar bar mass}

\section{Introduction} \label{sec:intro}

Decades of numerical studies have shown that bars can form naturally in dynamically cold stellar disks  \citep[e.g.,][]{Hohl1971,Ostriker+1973,Athanassoula02,Athanassoula03}. Observationally, dynamically cold disks appear to be a necessary, although not a sufficient, condition for bar formation, since some cold disks remain unbarred \citep{Sheth+12}. This intrinsic susceptibility of cold stellar disks to bar formation is reflected observationally in the high fraction of barred systems among disk galaxies, reaching approximately two-thirds in the local Universe \citep[e.g.,][]{deVaucouleurs+63,kmd+07,Buta+15}.
However, which physical processes regulate the growth of bars are still not fully understood.
Observational studies suggest that formation of bars follows the standard downsizing scenario, in which more massive galaxies formed  their bars earlier \citep{sheth+08,Kraljic+12}.
In this scenario, the bar has sufficient time to slow down and grow in length. This interpretation is consistent with observational studies showing that longer and stronger bars are preferentially found in quiescent galaxies  \citep{Geron21}, that bars are longer in more massive systems  \citep{Erwin+18,Kim21}, and that older bars are generally longer \citep{Kim14,Kim15,deSa-Freitas25}.

Studies point out that bars play a role in the secular evolution of their host galaxies. The formation and evolution of bars involve the exchange of angular momentum between the inner disk, the outer disk and dark matter halo \citep{Athanassoula03,Athanassoula+13}. Once formed, bars can drive gas toward the central regions through shocks along their leading edges \citep{Regan+1997,Regan1999,Sheth+05}, leading to central gas concentrations and enhanced nuclear star formation \citep{SersicPastoriza1967,Hawarden+1986,Ho1997,Sheth+05}. This process can build up a central gas reservoir \citep{Sakamoto1999,Sheth+05} and contribute to the formation and growth of central stellar structures, such as nuclear disks \citep{Bittner20,Gadotti2020,Schulthesis25}.
Observational studies have pointed to nuclear disks increasing in size and mass together with bar length \citep{Gadotti2025} and age \citep{deSa2025}, suggesting that these structures grow synchronously. 
However, these evolutionary processes may operate differently in lower-mass galaxies. 
In \citet{Verwilghen25}, using hydrodynamical simulations, they find that in low-mass systems the supernova-driven feedback can disrupt gas inflows, leading to clumpier bar morphologies and preventing the formation of stable central gas reservoirs.
Differences in bar properties linked to galaxy mass are observed across cosmic time.
Observations at low redshift (z$<1$) find that absolute bar length increases with galaxy stellar mass, especially for massive systems \citep[i.e., $\log($M$_\star/$M$_\odot)\sim10.1$,][]{Erwin+18,Kim21}, while JWST findings show constant values along galaxy stellar masses to high redshifts \citep{Leconte+26}. This epoch-dependent shift may suggest that bars formed with significant lengths regardless of their host galaxies' stellar masses in the early Universe, in agreement with simulations showing that bars forming at high redshift tend to be born longer than those forming at later epochs \citep{Fragkoudi+24,Lokas2025}.
However, these findings could also be driven by larger and more massive disks acquiring the conditions for bar formation earlier, resulting in an observed bar population biased toward larger structures.
Together, these results may point to bar growth pathways are not universal, but instead depend on galaxy mass, environment, and cosmic epoch. 

Bar growth is physically related to the trapping of new stellar orbits by bar resonances \citep{Athanassoula02,BeraldoeSilva2023}, progressively increasing stellar orbits supporting the bar. Thus, bar growth should also be gauged through the growth in stellar mass content of the bar structure. 
Although bar size and other structural properties have been used to trace these different evolutionary pathways, the role played by the stellar mass contained within the bar itself remains largely unexplored. Quantifying the stellar mass contained in bars therefore provides a physically motivated avenue to investigate how bar growth varies with galaxy mass and internal structure.

In this work we investigate how much stellar mass is contained within  bars in the local Universe, how the bar stellar mass varies with galaxy stellar mass and the presence of a central stellar component, and how these relations depend on other global galaxy properties. We aim to explore what these trends can reveal about the different pathways of bar growth.
We organize this paper in the following manner.
In Section \ref{Chapter2} we present our sample selection. In Section \ref{Chapter3} we explain our methodology for decomposing the light of each component and then convert the light to stellar mass. In Section \ref{Chapter4} we show our analysis and our main results. Section \ref{Chapter5} is our discussion. Section \ref{Chapter6} is the summary of our study.

\section{Sample and Data} 
\label{Chapter2}

We rely on deep mid-infrared observations at 3.6 and 4.5 $\mu m$ from the Spitzer Survey of Stellar Structure in Galaxies ($S^4G$; \citealt{Sheth+10,Munoz-Mateos+13}) to quantify the stellar mass in barred galaxies. The scatter in the mass-luminosity relation (M/L) is minimized in the 2-5 $\mu m$ range \citep{Driver2016}, making this spectral interval ideal for determining stellar masses. Studies demonstrate that the combination of 3.6 and 4.5 $\mu m$ observations provide a particularly effective way of mapping the underlying stellar mass distribution \citep[e.g.,][]{Meidt2014,Querejeta+15}, avoiding possible contamination of PAH emission (3.3 $\mu$m) within the 3.6 $\mu m$ band and of hot dust emission at 4.5 $\mu m$ \citep{Galliano2018}. 
Furthermore, studies point that near- and mid-IR bands are a good choice to easily identified bars \citep{Eskridge+00, kmd24, LeConte+24}.

$S^4G$ is the largest and deepest database with high-quality mid-IR images of nearby galaxies. 
Undertaken as part of the post-cryogenic phase of the Spitzer Space Telescope (SST; \citealt{Werner+04}), it consists of imaging a large sample of local galaxies at 3.6 and 4.5 $\mu$m with the Infrared Array Camera (IRAC; \citealt{Fazio+04}) down to a surface brightness $S* << 1$ M$_{\odot} / pc^2$. The $S^4G$ sample \citep{Sheth+10} consists of 2352 large ($D_{25} > 1^{\prime}$),  bright ($m_{Bcorr} < 15.5$, apparent brightness in the B-band, corrected for tilt, galactic extinction, and K correction) galaxies within the local volume ($d < 40$ Mpc), away from the Galactic plane ($|b| > 30 ^{\circ}$).

\label{Defining our sample}
Our study aims to analyze the distribution of stellar mass in nearby barred galaxies. 
To ensure a sample with well-defined bars, including those identified through visual inspection, we use the morphological classification from \citet{Buta+15}, based on visual inspection of $S^4G$ 3.6 $\mu$m band images. In addition to a detailed visual morphological classification, the authors categorized galaxies based on the presence of a bar. They applied the ``Family" criterion, assigning each galaxy a value from 0 to 1. Galaxies with $F\geq$0.5 are classified as barred, with $F=0.5$ corresponding to an SAB galaxy (intermediate bar) and $F=1$ indicating an SB galaxy (strong bar).
We select all $F\geq$0.5 galaxies, which results in a sample of 1201 barred galaxies.
Additionally, we considered the galaxy inclination and selected only galaxies with inclinations below 65$^{\circ}$.
This criterion is crucial for barred galaxies, since higher inclinations correspond to close-to-edge-on galaxies, for which the identification of a bar becomes unreliable. 
Considering these criteria, our sample is reduced to 524 barred galaxies. 

We also consider galaxies that receive a bar component in the decomposition in the 3.6 $\mu m$ band from \citet{Salo+15}. This 2D decomposition was performed for the entire $S^4G$ sample using GALFIT \citep{Peng+02,Peng+10}, where the final model includes between 1 and 4 components (i.e., disk, bulge, bar, nucleus, or secondary disk).  
They performed an analysis of the original data image and the final decomposition that resulted in quality flags for each galaxy (i.e., from 1 to 5, where Q=5 is the best quality decomposition). Galaxies that had a quality flag of 1 or 2 did not receive a final decomposition due to issues in the original data, such as very bright overlapping stars or strongly varying background noise. Other reasons for galaxies not receiving the best quality flag are being edge-on systems or very complex galaxies, for which a four-component model does not provide a good fit.  
Since we are interested in a sample of clear and well-characterized bars, considering both the quality of the original data and the decomposition results from \citet{Salo+15}, we select only galaxies with a quality flag equal to 5. We further focused on galaxies with a single disk component in their final model, excluding 12 galaxies with a secondary disk (i.e., outer disk with different slope) in their final fit.
For our mass determination analysis we rely on distance information from the S4G catalog and \citet{Bouquin+18} Table~1.

\begin{figure}[]
  \centering
  {\footnotesize
   \includegraphics[scale=0.5]{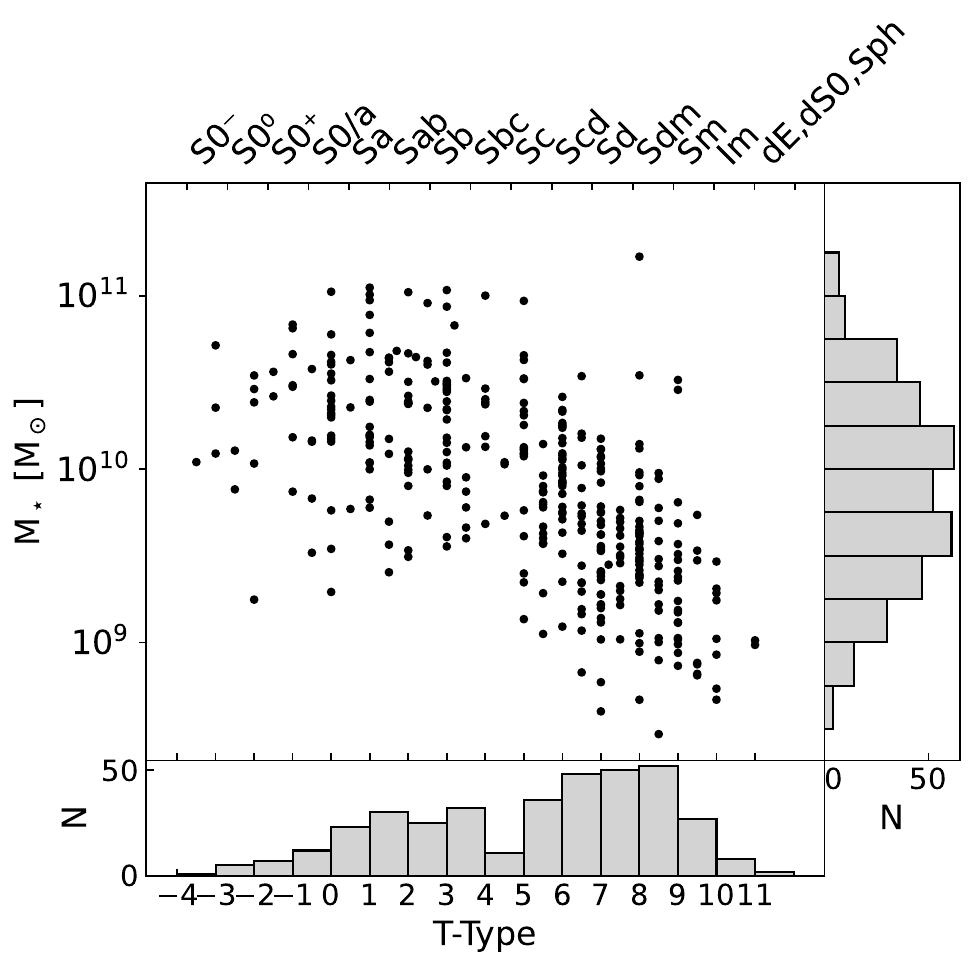}   
    \caption{The morphological types, masses and Hubble T-Types for our sample of 369 barred galaxies. The stellar mass is calculated using the 3.6 and 4.5 $\mu m$ fluxes, publicly available through $S^4G$ catalog, and the light to mass conversion from \citet{Querejeta+15}. The morphological types and Hubble T-Types are from \citet{Buta+15}.}
    \label{fig_mysample}   
    }
  \end{figure}

After applying all the criteria, our final sample consists of 371 barred galaxies. 
We perform an automated multi-band decomposition aiming to extend the final decomposition models obtained by \citet{Salo+15} also for the 4.5~$\mu$m band.
The use of both the 3.6 and 4.5~$\mu m$ bands is important for enabling a more accurate analysis of the stellar mass distribution across structural components, while minimizing potential contamination from PAH emission at 3.3~$\mu m$, as well as from star-forming regions and AGN activity. 
In Appendix \ref{Appendix_Underestimating} we demonstrate the need to extend the decomposition to the 4.5~$\mu m$ band for improved stellar mass estimates of individual components. 
We note that two galaxies in our sample, NGC 4035 and NGC 4105, did not have a mask image in 4.5 and 3.6~$\mu m$ band (respectively) included as part of the S4G Pipeline 1 and 2 products and were hence removed from our sample, reducing the latter to 369 galaxies.
Figure \ref{fig_mysample} shows the stellar mass and morphological type distribution of our sample.
For our analysis, we used the 3.6 and 4.5 $\mu m$ images and the masks for foreground stars provided by the $S^4G$ collaboration in the form of Pipelines 1 and 2 products, respectively. 
All the data used in this work are publicly available through the IRSA $S^4G$ webpage\footnote{Available at: \url{http://irsa.ipac.caltech.edu/data/SPITZER/S4G}}.

\section{Galaxy image decomposition}
\label{Chapter3}
We aim to isolate the 3.6 and 4.5 $\mu m$ light contribution of each stellar structure in our galaxy sample.
With 2D decomposition, we create a model that describes the surface brightness profile for each stellar structure using GALFITM \citep{Bamford2011,Vika+14,Haussler+22}, which is based on GALFIT \citep{Peng+02,Peng+10}, improved to fit several bands simultaneously.
As GALFIT, GALFITM performs a non-linear least squares fit by minimizing residuals in the 2D plane.
The final model is defined as the sum of the images of the individual galaxy components.
Once we have the distinct light contribution of each stellar component in these mid-IR bands, we can convert the light to stellar mass, as we detail below.

\subsection{Fitting stellar components with GALFITM -- models with varying complexities}
\label{equationsGALFITM}

GALFITM \citep{Bamford2011,Vika+14,Haussler+22} is an extension of GALFIT \citep{Peng+02,Peng+10}, designed to fit analytical functions to the surface brightness profiles of galaxies across multiple bands simultaneously.
The fitting is carried out using a non-linear least-squares optimization via the Levenberg-Marquardt algorithm, which minimizes the residuals between the model and the observed data. The fit is iteratively refined until the chi-squared ($\chi^2$) value converges, i.e., shows no significant improvement. The final fit quality is assessed using the reduced chi-squared ($\chi^2_{\nu}$) statistic

\begin{equation}
    \chi^2_{\nu} = \frac{1}{N_{dof}} \sum_x \sum_y \frac{[\, Data(x,y) - Model(x,y) \,]^2}{\sigma(x,y)^2} \vspace{0.5cm},
\end{equation}

\noindent
where $N_{dof}$ is the number of degrees of freedom. 
The final decomposition models for our sample consist of 2 to 4 components. The simplest models include disk and bar components, while the most complex models incorporate disk, bar, bulge and nucleus. 
Each component in the final model is fitted using the following analytical functions:

\begin{enumerate}

    \item Bulge - Sérsic profile \newline To fit the bulge light, we adopt a Sérsic profile, given by:
    
\begin{equation}
     {\textstyle\sum(r)} = {\textstyle\sum}_e \exp\left(-\kappa \left[(r/R_{e})^{1/n} - 1\right]\right) \vspace{0.5cm},
\end{equation}

where n is the Sérsic index, $\Sigma_e$ the surface brightness within the effective radius\footnote{the isophotal radius encompassing half of the total flux of the component.} $R_e$, and $\kappa$, a normalization constant determined by $n$. 
The Sérsic index $n$ describes the shape of the radial profile, becoming steeper as $n$ increases; $n$ = 1 corresponds to an exponential profile, and $n$ = 4 corresponds to a de Vaucouleurs profile. 
For this component, the parameters in GALFITM are: $x_c$ and $y_c$ (center position in pixel values), total magnitude, effective radius, Sérsic index, axis ratio (b/a), and position angle.
For effective radius, axis ratio, and position angle, we fit the same value in both bands (i.e., the parameters are fit but with the same value in both bands). For the total magnitude and the Sérsic index, we allow for different fits in each band.

    \item Disk - exponential profile \newline To fit the disk light, we use an exponential profile, given by:

\begin{equation}
    \textstyle\sum(r) = {\textstyle\sum}_0 q^{-1} exp\left( -r/h_r \right),
\end{equation}

where the parameter $h_r$ denotes the disk scale length, and $q = cos (i)$, where i represents the inclination of the disk. The term ${\Sigma}_0 q^{-1}$ corresponds to the central surface brightness projected onto the plane of the sky, assuming no extinction.
For this component the parameters in GALFITM are $x_c$ and $y_c$ (center position in pixel values), total magnitude, disk scale length, axis ratio (b/a), and position angle.
We fix the center, axis ratio and position angle. For the disk scale length, we fit the same value in both bands. For the total magnitude, we allow different fits in each band.

    \item Bar - modified Ferrer profile \newline Bars can be modeled with the modified Ferrer profile which is given by:

\begin{equation}
    \textstyle\sum(r) = 
    \left \{
\begin{array}{lc}
{\textstyle\sum}_0 \left[1 - (r/r_{out})^{2 - \beta} \right]^{\alpha}  &, \,\,\,\, r < r_{out} \\
0 &, \,\,\,\, r \geq r_{out}\\
\end{array}
\right.
\label{ferrerfunction}
\end{equation}

where $r_{out}$ defines the outer truncation radius of the profile (i.e.,the region where the flux comes to zero), $\alpha$ determines the sharpness of this truncation, $\beta$ defines the central slope of the profile, and ${\Sigma}_0$ represents the central surface brightness in the plane of the sky.
For this component, the parameters in GALFITM are: $x_c$ and $y_c$ (center position in pixel values), central surface brightness, outer truncation radius, $\alpha$ (outer truncation sharpness, fixed to a value of 2), $\beta$ (central slope, fixed to the value of 0), axis ratio ($b/a$), and position angle.
We fix the center, as well as the $\alpha$ and $\beta$ values. For the outer truncation radius, axis ratio, and position angle, we fit the same value in both bands. For the central surface brightness, we allow for different fits in each band.

    \item Nucleus - PSF

A nucleus is fit when the galaxy contains an unresolved central component. The fit is based on the PSF used.
For this component, the parameters in GALFITM are $x_c$ and $y_c$ (center position in pixel values) and the total magnitude. The free parameter is the total magnitude, $m_{psf}$ with a different fit in each band, while we use the fix values for position x and y. 
    
\end{enumerate}

For the decomposition to converge to a final model, it is important to choose good initial parameters (e.g., center coordinates, zero-point magnitudes). 
We build upon the decomposition performed by \citet{Salo+15} in the 3.6 $\mu$m band for the $S^4G$ sample, using its output parameters as initial values for our multi-band decomposition. We extend this approach to the 4.5 $\mu$m band, allowing us to combine the information from both bands for a more robust estimate of the stellar mass distribution.
First, we perform a single Sérsic fit to each galaxy to estimate its center coordinates.
We then fix the center coordinates, $x_c$ and $y_c$, for all components to these values, ensuring that all components share a common center consistent with the estimated galaxy center.
We estimate the zero-points for bands 3.6 and 4.5 $\mu m$ following the instructions in the IRAC Instrument Handbook, where the zero magnitude flux density is 280.9 Jy and 179.7 Jy in the 3.6 and 4.5 $\mu m$ bands, respectively. The units of the images in $S^4G$ catalog are in MJy/sr, and the pixel scale is 0.75$^{\prime\prime}$/pix. To convert MJy/sr to Jy/pixel, we adopt:
1 [MJy/sr] = 13.22 $\times 10^{-6}$ [Jy/pixel]. After converting to AB magnitudes, we obtain the zero-point value of 21.097 for both bands, consistent with the zero-point values used in \citet{Watkins2022}.
For our final decomposition models, we choose to use the same components from \citet{Salo+15} used in their final models. This approach gives us four possible models for our sample:
\begin{itemize}
    \item \textbf{DBARBN}: Disk + Bar + Bulge + Nuclear Point Source

    \item \textbf{DBARB}: Disk + Bar + Bulge 

    \item \textbf{DBARN}: Disk + Bar + Nuclear Point Source

    \item \textbf{DBAR}: Disk + Bar 
\end{itemize}

In Appendix \ref{preparing ingredients}, we describe in detail the preparation of data to perform  our decomposition.

\subsection{Executing the decomposition -- dataset of final models for our barred sample}

To perform the automated decomposition, we first perform single-Sérsic fits for our entire sample in both bands. This initial step serves to evaluate the quality of our images, ensuring that the decomposition is feasible across the entire sample. It also allows us to estimate the central coordinates of each galaxy. While the center is a free parameter in the single-Sérsic fits, we fix it for multi-component models. This is necessary because, if left free, the different components may converge to different center positions, potentially resulting in an inconsistent or unreliable final model. A well-estimated center is also crucial for obtaining reliable fits for central components, such as the bulge and the nuclear point source.
In our sample of 369 barred galaxies, we obtain single-Sérsic fits for 355 galaxies. For the remaining 14 galaxies, we perform the multi-component decomposition considering the central pixel as the center of the galaxy -- a reasonable approximation, since we cropped the images using the center of the galaxy provided in the image headers.

After the single Sérsic fit, we perform the fit of the final models. We adopt the same final models from \citet{Salo+15}. We obtain successful fits for 344 out of the 369 galaxies. The 25 galaxies for which we do not obtain satisfactory fits are removed from our final analysis. Of these 25 galaxies, 11 are classified as unreliable fits (i.e., flagged with an asterisk (*) by GALFITM) and 14 as unsuccessful fits (i.e., GALFITM can not fit a final model). Table \ref{Table2} summarizes the distribution of final fit models used for our sample. 
The inability to obtain a reliable final model for some galaxies appears to be related to the image quality in one or both bands. Although we select galaxies classified with quality flag 5 (i.e., the highest image quality; \citealt{Salo+15}), in a few cases some galaxies exhibit image artifacts that may affect the model fitting. In other cases, the galaxy appears too faint, making the fit unreliable in at least one band. This results in flagged values (marked with an asterisk) in the output, such as unrealistically faint magnitudes for the nuclear component, excessively high Sérsic index for the bulge, or very low axis ratios for the bar component.

\begin{table}[t]
\centering
{\footnotesize
\begin{tabular}{cccccc}
\hline
Final model  & $N_{gal}$ & \multicolumn{4}{c}{Mass ratio [\%]}\\ 
(initial sample: 369) & & Disk & Bar & Bulge & Nucleus \\ \hline
DBARBN       & 16 & 68.5 & 11.9 & 14.9 & 3.0  \\
DBARB        &  87 & 70.1 & 10.7 & 15.4 & -   \\
DBARN        & 75 & 92.8 & 5.4 & - & 0.7 \\
DBAR         & 166 & 93.2 & 6.8 & - & -    \\ \hline
Total sample \\(successful fits)        & 344 \\ 
Unsuccessful fits & 14 \\
Unreliable fits  & 11 \\
\hline
\end{tabular}
\caption{The distribution of final models obtained from our multi-band decomposition. The stellar mass ratio columns shows the median stellar mass of each component with respect to the total stellar mass for each final model. The median mass fractions were computed independently for each component, therefore, their sum is not necessarily equal to 100\%.}

\label{Table2}
}
\end{table}

\begin{figure*}[ht]
  \centering
  {\footnotesize
   \includegraphics[width=1.0\textwidth]{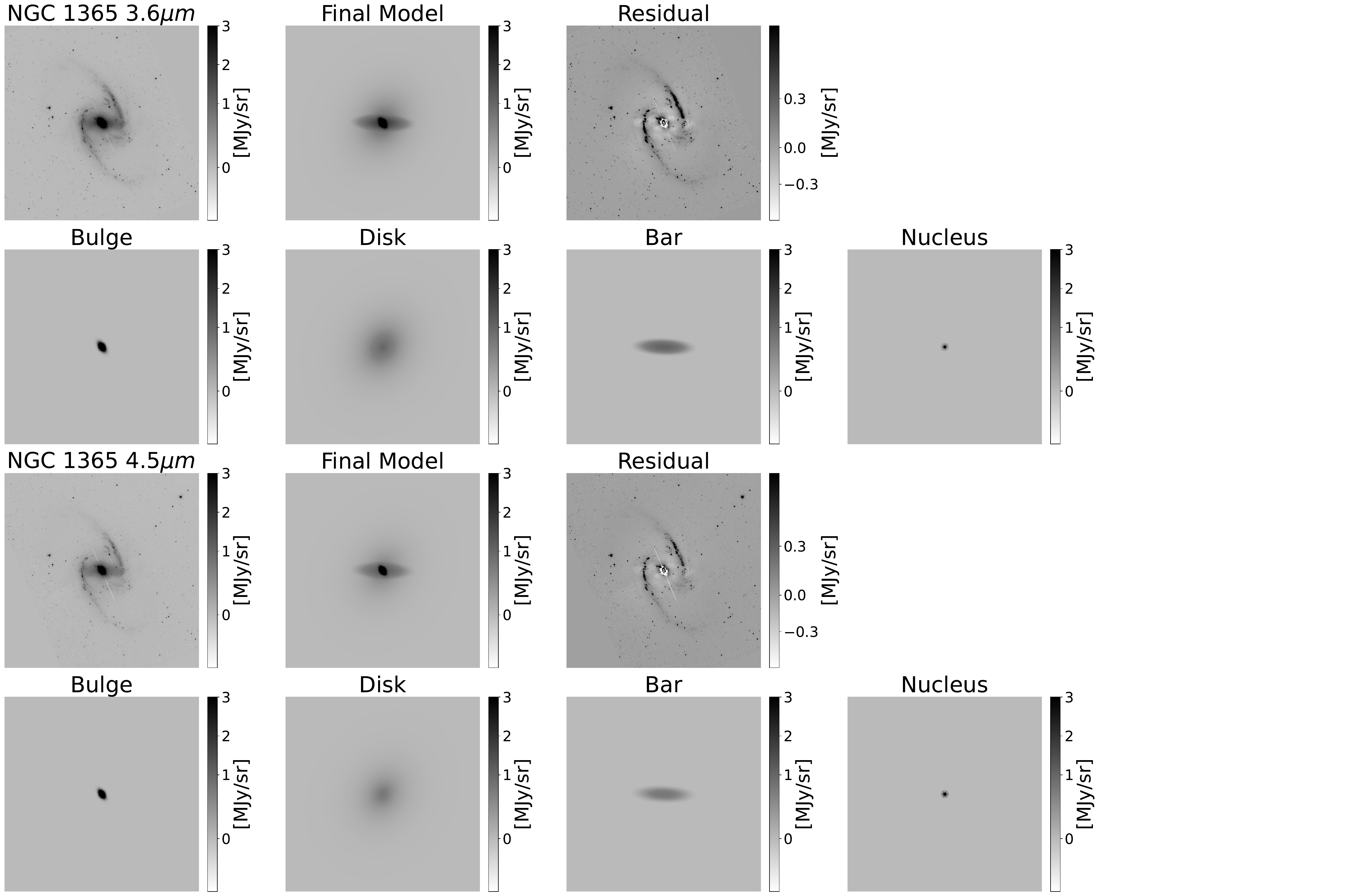}
    \caption{Four-component decomposition of the galaxy NGC 1365. The first and third rows show the galaxy image (left), the final decomposition model (center), and the residual (right) in the 3.6 and 4.5 $\mu m$ bands, respectively. The second and fourth rows display the individual components that make up the final decomposition model—bulge, disk, bar, and nucleus—from left to right in the corresponding bands.}
    \label{fig_ex_gal_model_res_comps}   
    }
  \end{figure*}

The output from GALFITM provides final model images with all components summed, as well as residual images (i.e., galaxy image minus the model). 
Figure \ref{fig_ex_gal_model_res_comps} shows an example of the galaxy image, the final model, and the residual, along with the separated images of each component contributing to the final model. This allows us to quantify the light contribution of each component to the final decomposition model using the following relations: the sum of the pixel values of all individual component images for a given galaxy equals the final model image; the fractional light contribution of each component is given by the ratio between the component image and the final model image.
Using this approach, we assess the light contribution of each structural component in the different decomposition models across our sample.

\subsection{Converting mid-infrared flux to stellar mass}

Knowing the fluxes in the 3.6 and 4.5 $\mu$m bands, we use the conversion proposed \citet{Querejeta+15}. This study used more than 1600 galaxies from the $S^4G$ sample to derive stellar mass maps, employing a method that separates the dominant light from old stars from dust emission, which can contribute to the 3.6 $\mu$m flux. Based on this analysis, an effective mass-to-light ratio (M/L) was calibrated, leading to the following equivalent equation

\begin{equation}
    \frac{M_{\star}}{M_{\odot}} = 10^{8.35} \times \left(\frac{F_{3.6}}{Jy} \right)^{1.85} \times \left(\frac{F_{4.5}}{Jy}\right)^{-0.85} \times \left( \frac{D}{Mpc} \right)^2
    \label{equacaomassa}
\end{equation} 

\noindent
where $M_{\star}$ is the stellar mass in solar masses, $F_{3.6}$ and $F_{4.5}$ are the fluxes in the 3.6 and 4.5 $\mu$m bands in Jansky (Jy), and $D$ is the galaxy distance in Mpc.
Our approach to study the stellar mass distribution is based on our galaxy decomposition, whose output provides a two-dimensional model representing the light distribution in both bands for each stellar structure, which together account for the total light of the final model. Using the output images from GALFITM, we quantify the fraction of light associated with each stellar component (e.g., the stellar disk accounting for $\sim$80$\%$ of the total model light).
The final model total flux is the sum of the fluxes associated with the components that represent the galaxy in the final model.
For the distances, we use the values by \citet{Bouquin+18}, their Table~1, who provide a catalog with updated distance estimated for many $S^4G$ galaxies. When these values are not available, we use the ``dmean" column from the $S^4G$ catalog, available on the IRSA website.
The fluxes of our final models do not reflect residual structures (e.g., spiral arms, shells) that may be present. In spite of this, when comparing our final model fluxes with the observed fluxes reported in \citet{Munoz_Mateos+15}, we find a general trend of final models with total fluxes above these observed fluxes in both the 3.6 and 4.5 $\mu m$ bands. We attribute this difference to the fact that GALFITM models the light distribution smoothly toward the outer regions, resulting in an additional light contribution beyond the isophotal limit used to measure the observed flux, which is truncated at a surface brightness level of 25~AB~mag~arcsec$^{-2}$ (see \citealt{Munoz_Mateos+15} for details). We correct for this discrepancy by defining a total ``decomposed" flux:
\begin{equation}
    F_{\rm decomp} = F_{\rm model} + F_{\rm residual}, \label{eq_Fdecomp}
\end{equation}
where the residual flux ($F_{\rm residual}$) is integrated into the total flux.
We then derive an equivalent flux for each component as
\begin{equation}
    F_{\rm comp} = \frac{F_{\rm galfitcomp}}{F_{\rm decomp}} F_{\rm obs}, \label{Fcomp}
\end{equation}
thereby preserving the relative contribution of each component ($F_{\rm galfitcomp}/ F_{\rm decomp}$) while ensuring that their summed flux is consistent with the observed flux ($F_{\rm obs}$).
In Appendix \ref{Comparison galaxy and model stellar mass} we show that the stellar mass derived for our galaxies using this corrected flux is in agreement with that derived from the 3.6 and 4.5 $\mu m$ magnitudes reported in \citet{Munoz_Mateos+15}\footnote{Data available at: \href{https://irsa.ipac.caltech.edu/data/SPITZER/S4G/overview.html}{https://irsa.ipac.caltech.edu/data/SPITZER/S4G}.}.

\begin{figure}[]
  \centering
  {\footnotesize
   \includegraphics[width=0.48\textwidth]{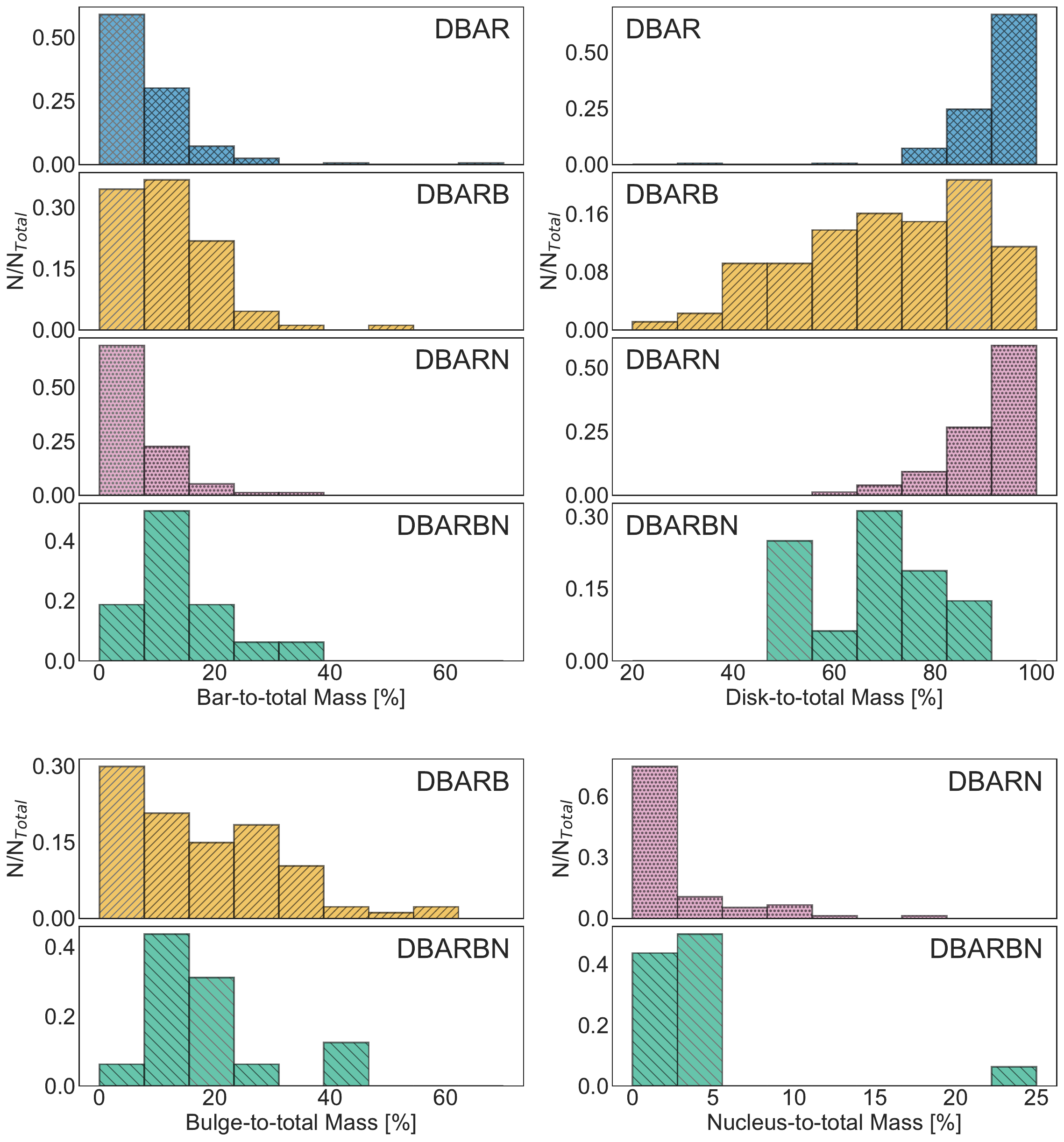}
    \caption{The component-to-total stellar mass distribution for the different component present in the final models of our sample. Each image shows the relative stellar mass for a different component. The different colors shows our final models. }
    \label{fig_mass_distribution}   
    }
  \end{figure}

With our decomposition in hand, we convert the light of each component into stellar mass. Figure \ref{fig_mass_distribution} shows histograms of the stellar mass distribution for the different final decomposition models. Our decomposition results in a catalog containing the stellar mass associated with each structural component, as derived from our final models. Table \ref{table3} summarizes our mass distribution results based on the final models, including the total stellar mass in each model, the stellar mass associated with each structural component, and the component-to-total mass ratios, which quantify the fraction of mass contained in each component relative to the total model mass.

\begin{table}
\centering
\setlength{\tabcolsep}{4pt}

\fontsize{7}{8}\selectfont


\begin{tabular}{ccccccccccc}
\toprule

Galaxy & Model & Total &  \multicolumn{4}{c}{Mass ratio [\%]}  \\
 &  & log(M$_\star/$M$_\odot$) & Disk  & Bar  & Bulge  & Nucleus \\
 
\midrule
ESO012-010 & DBAR & 9.5 & 94.5 & 5.5 & -- & -- \\
ESO013-016 & DBAR & 9.5 & 95.6 & 4.4 & -- & -- \\
ESO026-001 & DBARN & 9.4 & 91.6 & 8.1 & -- & 0.3 \\
ESO027-001 & DBARB & 10.0 & 84.2 & 8.4 & 7.4 & -- \\
ESO048-017 & DBARN & 8.9 & 88.2 & 11.5 & -- & 0.4 \\
ESO079-007 & DBAR & 9.4 & 96.2 & 3.8 & -- & -- \\
ESO238-018 & DBAR & 9.4 & 96.2 & 3.8 & -- & -- \\
ESO287-037 & DBAR & 9.7 & 94.5 & 5.5 & -- & -- \\
ESO340-017 & DBAR & 9.6 & 83.0 & 17.0 & -- & -- \\
ESO341-032 & DBAR & 9.7 & 96.4 & 3.6 & -- & -- \\
ESO357-012 & DBAR & 9.2 & 94.9 & 5.1 & -- & -- \\
ESO408-012 & DBARN & 9.5 & 97.1 & 2.7 & -- & 0.3 \\
\hline
\end{tabular}
\caption{First rows of the table with the stellar mass distribution ratio for each galaxy.
The complete table is available at \href{https://github.com/Cavalcante-Coelho/Stellar-Mass-Census-of-Local-Bars}{https://github.com/Cavalcante-Coelho+26}.}
\label{table3}

\end{table}

\section{Analysis and Results}
\label{Chapter4}

\subsection{Sérsic index of galaxies with a bulge component}

In our sample of 344 barred galaxies, 103 include a bulge component in their final decomposition model. But what is the nature of these bulges? 
Over the past two decades, studies have increasingly focused on the distinction between classical bulges and nuclear disks, also commonly referred to as pseudo-bulges \citep{KK+04,Gadotti2009}. 
Unlike classical bulges, nuclear disks are stellar structures characterized by rapid rotation 
and low velocity dispersion, and have nearly circular orbits 
\citep{Gadotti2020}. 
Considering the role of bars in the secular evolution of galaxies, bar-driven gas inflow toward the central regions appears to be linked to the formation and growth of nuclear disks \citep{Seo19,Gadotti2020,Schulthesis25,Kwak+26}.
Therefore, the central structures referred to as bulge components in this paper may correspond to nuclear disks or other central stellar structures that can be described by a Sérsic profile.

\begin{figure}[ht]
  \centering
  {\footnotesize
   \includegraphics[width=0.45\textwidth]{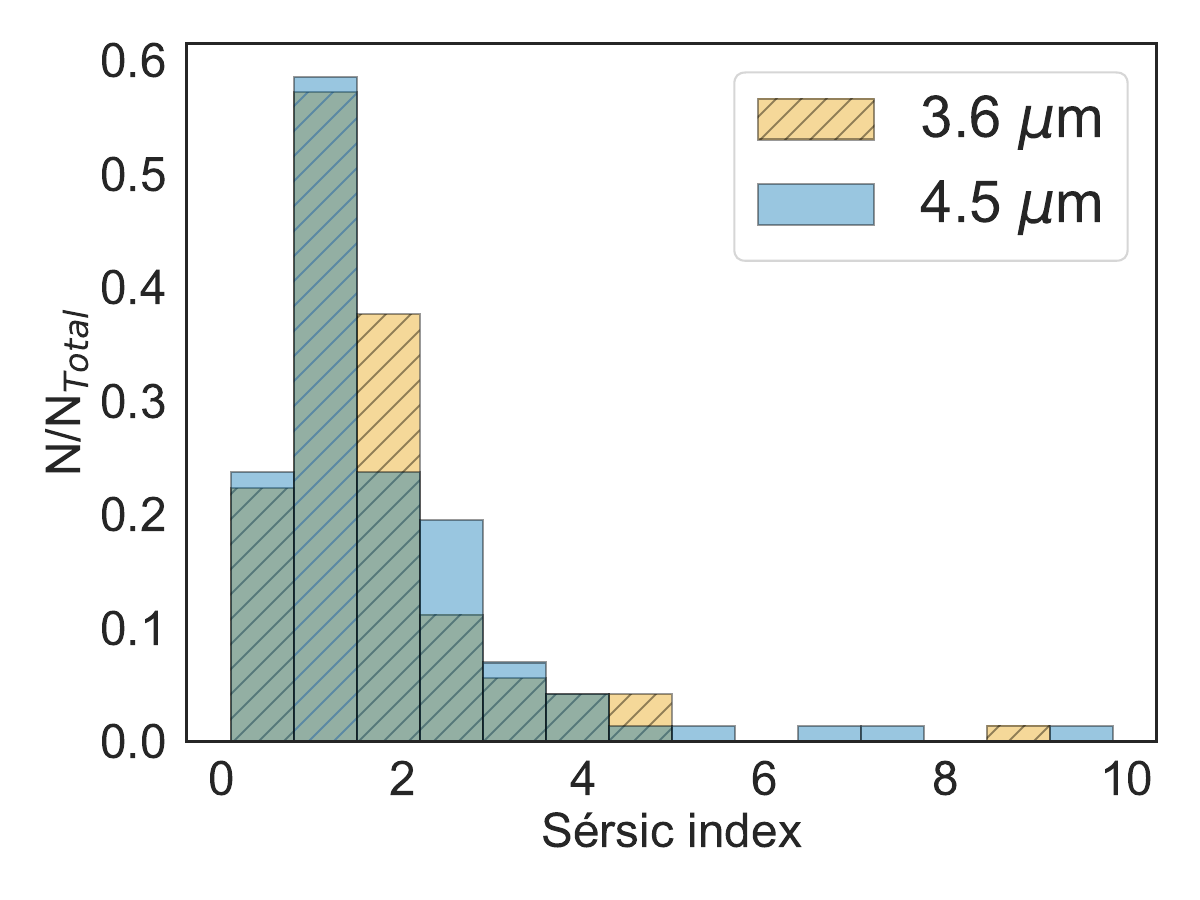}
    \caption{The Sérsic index values for the bulge components for 3.6 $\mu$m band (yellow, dashed) and 4.5 $\mu$m (blue, solid). Both distributions have median values of 1.38.}
    \label{sersic_index}   
    }
  \end{figure}

To identify possible nuclear disks in our sample from a photometric perspective using the \citet{Sersic+63} function (i.e., bulge Sérsic index n $\leq$ 2; \citealt{Gadotti2009,Gadotti2020}), we examine the Sérsic indices in both bands for galaxies that include a bulge component.
Figure \ref{sersic_index} shows the Sérsic index distribution in our sample for final models that include a bulge component (i.e., DBARBN and DBARB) in both 3.6 and 4.5 $\mu m$ bands, with median values of 1.38 for both bands. In general, we find the Sérsic index values n $\leq$ 2 for most galaxies, with 76 systems (74\%) in 3.6 $\mu m$ band and 71 systems (69\%) in 4.5 $\mu m$ band. This may indicate that nuclear disks are common -- and perhaps even expected -- in barred galaxies with central structures that are photometrically similar to bulges.

\citet{Gadotti2020} present a detailed kinematic analysis of galaxies hosting nuclear disks that are part of the TIMER sample \citep{TIMER2019}.
Our sample overlaps with the TIMER sample for 12 galaxies.
We find that 10 out of the 12 galaxies have n $\leq$ 2 when considering only the 3.6 $\mu m$ band, while 8 out of 12 galaxies meet this criterion in both bands.
One galaxy with a verified nuclear disk and with n $>$ 2 is NGC 1291. This galaxy hosts a nuclear disk and an inner bar \citep{Gadotti2020}, but in \citet{Mendez-Abreu19} also shows that it contains a small classical bulge. 
The comparison between our results, based on photometrically determined Sérsic indices, and the kinematic analysis of \citet{Gadotti2020} supports our interpretation that the central structures identified in our work with n $\leq$ 2 are very likely nuclear disks. However, a definitive confirmation naturally requires kinematic data, which are currently not available for our full sample and is beyond the scope of the current work.

\subsection{Impact of bulge on bar size and mass}

\begin{figure*}[ht]
  \centering
  {\footnotesize
   \includegraphics[width=0.9\textwidth]{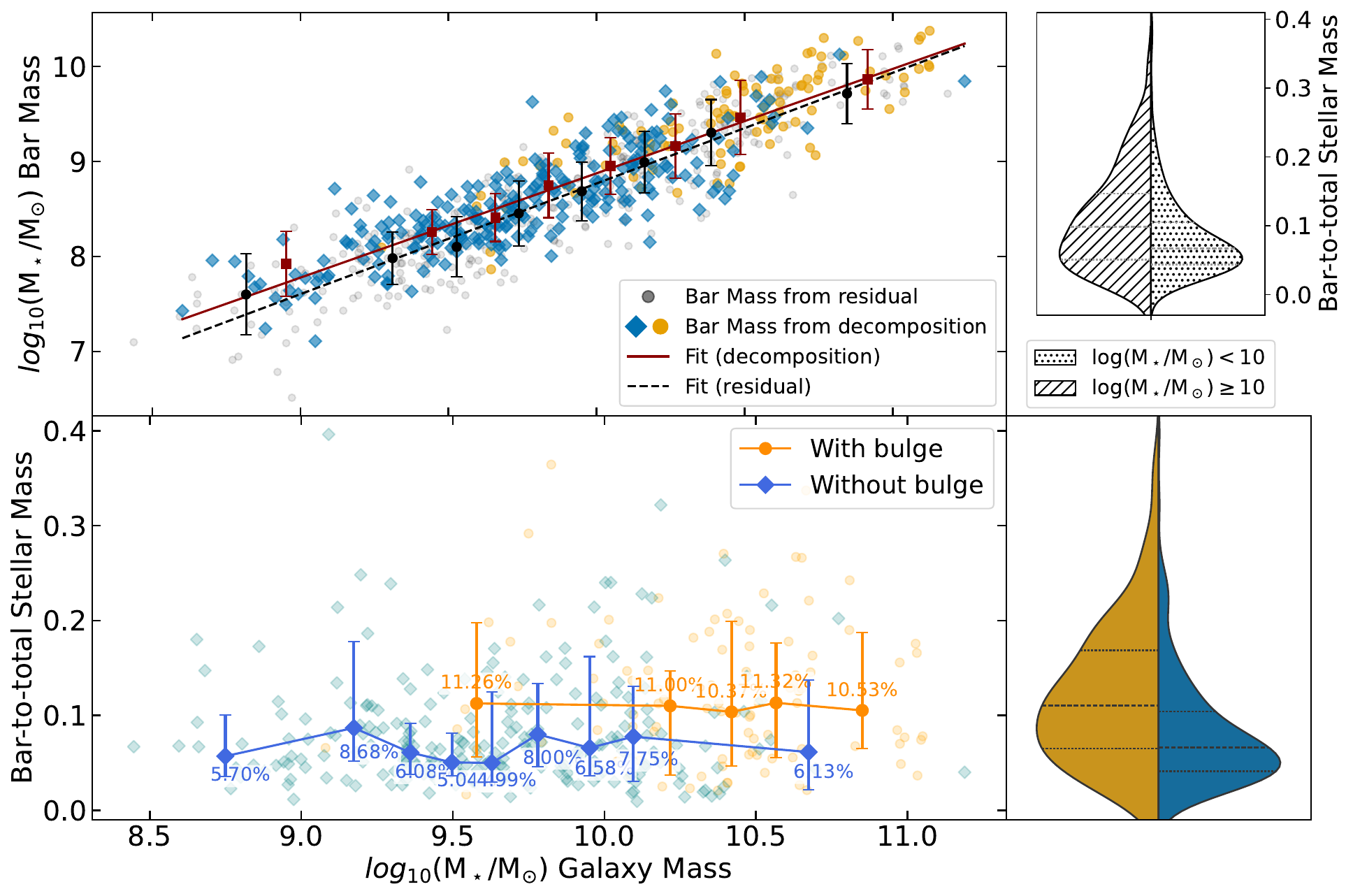}
    \caption{The relation between the total stellar mass and the stellar mass contained in the bar component (top left), comparison of the bar-to-total stellar mass between high- and low-mass galaxies (top right) and a comparison between bar-to-total stellar mass for galaxies with and without bulge component (bottom) for different total stellar mass. In the top panel the red squares indicate the median values of bar stellar mass considering the light from Ferrer function, while the black circles are the same but for second approach, using residual and bar light. Each bin has 43 galaxies for the both cases. We fit a linear relation, using curve\_fit from PYTHON scipy.optimize package; Jones et al. 2001. In the bottom panel the orange circles are the median values for galaxies with bulge (103 galaxies) in its final model, while the blue diamond are for galaxies without bulge (241 galaxies). Each bin has 20-21 and 26-27 galaxies for the sub-samples with and without bulge, respectively. The error bars correspond to the 16th and 84th percentiles, encompassing 68\% of the distribution in each bin. }
    \label{fig_relation_mass_bar_model}   
    }
  \end{figure*}

We find a trend to more massive galaxies with more massive bars, illustrated in Figure \ref{fig_relation_mass_bar_model}.
As massive galaxies tend to host longer bars \citep{Erwin+18,Kim21}, this is a natural result.
To probe whether this trend extends to the bar mass content relative to the total galaxy stellar mass (i.e., to the bar-to-total stellar mass ratio) 
we perform a hypothesis test to determine whether the low- and high-stellar-mass subsamples show significantly different distributions.
We apply the Wilcoxon–Mann–Whitney test (U test), a non-parametric hypothesis test that assesses whether two independent samples originate from the same parent distribution.
We consider the bar-to-total stellar mass ratio distributions of two similarly-sized sub-samples of low- and high-masses (top right panel on Figure \ref{fig_relation_mass_bar_model}): a lower-mass galaxy sample ($M_\star < 10^{10}$ M$_{\odot}$) comprised of 200 galaxies, with a median bar-to-total stellar mass ratio of 6.7\%; and a high-mass galaxy subsample ($M_\star \geq 10^{10}$ M$_{\odot}$; sample size 144 galaxies), with a median value of 9.9\%. The p-value obtained for this test is $ \sim 5 \times 10^{-4}$, which is below the 0.05 threshold (significance level adopted). 
This result suggest that massive galaxies host bars with higher relative stellar mass than their low-mass counterpart.

Considering that galaxies whose final decomposition models include a bulge component, i.e., DBARBN (disk-bar-bulge-nucleus) and DBARB (disk-bar-bulge) models, tend to be those of higher galaxy masses (mass range; see Figure~\ref{fig_relation_mass_bar_model}), bulge prominence appears to be linked to higher bar masses. To assess whether bar-to-total stellar mass ratios in galaxies with and without bulges in their final models present statistically significant differences, we perform the hypothesis test. 
We separate two samples considering their final model: a non-bulge sample (final models DBAR and DBARN; sample size 241 galaxies; median value 6.6\%), and a bulge sample (final models DBARB and DBARBN; sample size 103 galaxies; median value 11.3\%). The p-value obtained from the hypothesis test for these two samples is $ \sim$8$ \times 10^{-8}$, which is also below 0.05.
This result point to a possible connection between more massive bars and more massive galaxies hosting bulge components.

In our decomposition, we use the modified Ferrer function (Equation \ref{ferrerfunction}) to fit the brightness distribution in the bar region. To avoid introducing an excessive number of free parameters, we fixed $\alpha$ and $\beta$ in the modified Ferrer profile used to model the bar.
The Ferrer function assumes a flat surface-brightness profile for the bar; such have been shown to be characteristic of bars in more massive galaxies (e.g., \citealt{Kim15,Erwin2023}). In lower-mass galaxies, however, bar surface-brightness profiles tend to show a more continuous decline, without the extended flat region characteristic of bars in more massive galaxies \citep{Kim15}. 
Considering that bars in low-mass galaxies may not be as adequately represented by a Ferrer model, we verify the robustness of our findings across all masses by applying an alternative approach to estimate the stellar mass contained within the bar region.
This approach consists of assessing the residual light in the bar region when fitting the original galaxy image using only the other components included in the galaxy's final model (disk, bulge and/or nucleus), without explicitly fitting a bar component (see Appendix \ref{appendix_barwithresidual} for details).
Briefly stated, we define the bar region inside its length (based on both Ferrer truncation radius and bar semi-major axis from \citealt{herrera-endoqui2015}), position angle, and ellipticity. 
From the light enclosed within this region in both bands, we calculate the bar stellar mass without explicitly fitting a bar component.
Figure \ref{fig_relation_mass_bar_model} 
shows the bar mass estimates based on our decomposition and this alternative calculation based on image residuals (i.e., science image - disk model - bulge model - nucleus model). 
The residual-based approach yields lower bar stellar masses, particularly for low-mass galaxies, but preserves the same overall trend.

\begin{figure*}[ht]
  \centering
  {\footnotesize
   \includegraphics[width=0.9\textwidth]{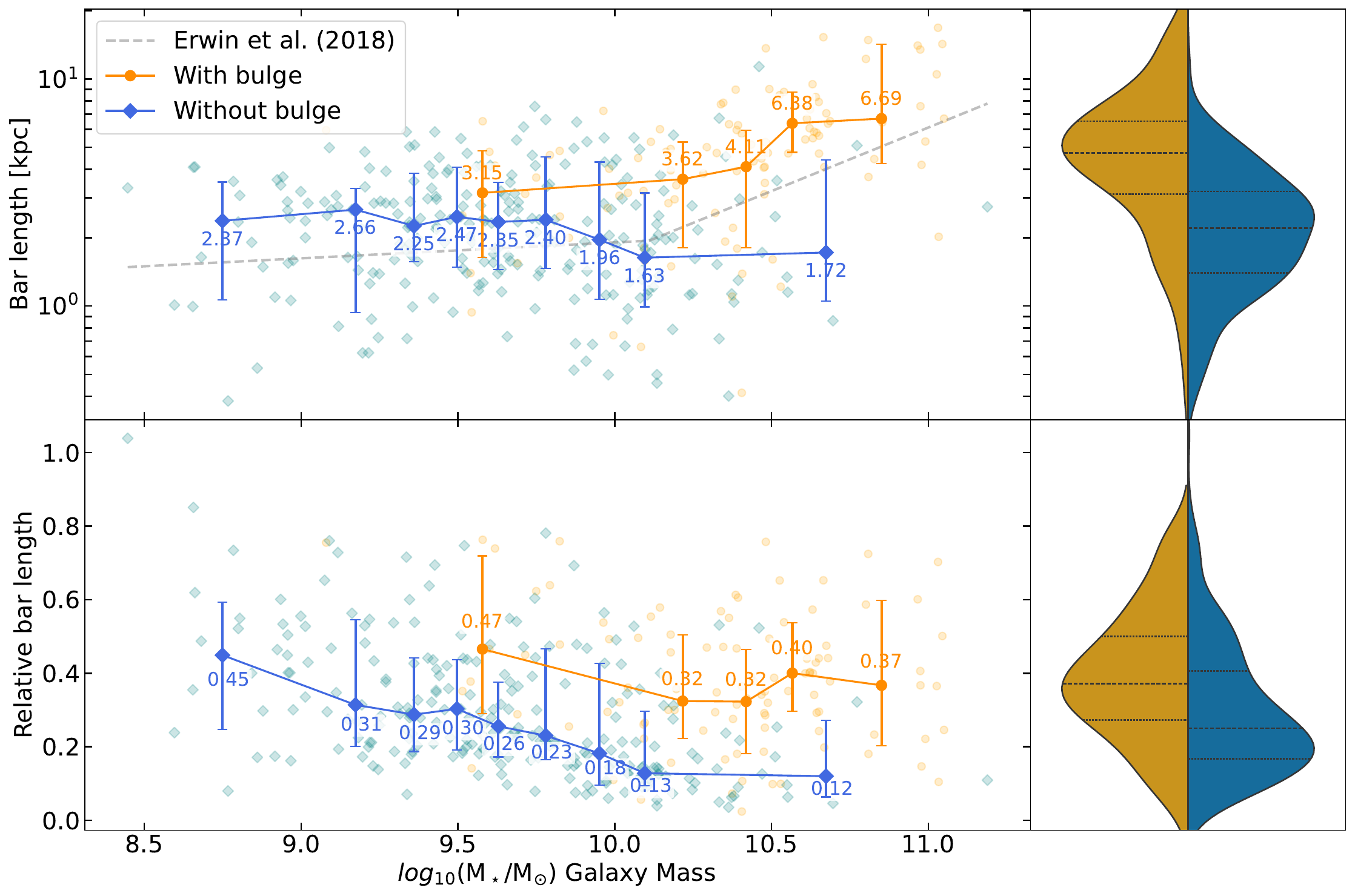}
    \caption{Relation between bar length (bar semi-major axis) and total galaxy stellar mass, where the median in galaxy mass bins are highlighted and connected. The top panel shows how the bar length varies with galaxy mass for the entire sample, the middle panel shows the length difference considering the bulge in the final model, and the bottom panel shows the relative bar length (bar sma/galaxy R$_{25.5}$). The bar length is the truncation radius obtained from the Ferrer function fit applied to the galaxies in our sample, as described in Section \ref{equationsGALFITM}. The output value is converted to kiloparsecs. In the top panel, the pink circles are the median values of the length from our decomposition, and the small gray circles are the median values considering \citet{herrera-endoqui2015} bar length values for our sample. Each bin has 43 galaxies.
    The dashed line is the relation found in \citet{Erwin+18}, for the author's main spiral sample. In the middle and the bottom panels, the orange circle are the bar length median values from galaxies with bulge in its final model (103 galaxies), while the blue diamond represents galaxies without bulge in its final model (241 galaxies). Each bin has 20-21 and 26-27 galaxies for the sub-samples with and without bulge, respectively. The error bars correspond to the 16th and 84th percentiles, encompassing 68\% of the distribution in each bin.}
    \label{fig_relation_size_bar_mass_total}  
    }
  \end{figure*}

The linear relation that we find between bar stellar mass and total galaxy mass differs somewhat with what has been observed when considering bar size, as discussed by \citet{Erwin+18}.
In his work, the author points to a broken power law, where a positive correlation between absolute bar size and total stellar mass is identified only for galaxies with $\log_{10}(M_\star/M_\odot)>10.1$, while for lower-mass galaxies the absolute bar size remains nearly constant regardless of host galaxy mass. 
Other studies also find a similar trend \citep{Diaz-Garcia15,Wang25,Hutchinson-Smith26}. By adopting our GALFITM bar truncation radius as a proxy for the bar length, we verify in Figure~\ref{fig_relation_size_bar_mass_total} (top panel) that we also recover the broken power-law presented by \citet{Erwin+18}. We note that although the true bar sizes need to be deprojected, we reproduce this result based on projected values.

To assess whether limited spatial resolution could account for the bar lengths observed in low-mass galaxies, we assume that bar semi-major axes can be reliably measured only when they exceed twice the PSF FWHM \citep{Guo+24}. For the $S^4G$ 3.6 $\mu$m PSF FWHM of 1.89$^{\prime\prime}$, and considering the criterion of at least 3 $\times$ PSF FWHM, this corresponds to a minimum measurable bar length of approximately 0.69 kpc at the mean distance of our sample, $\sim$25.2 Mpc. Adopting the maximum sample distance of $\sim$40 Mpc gives a more conservative limit of approximately 1.1 kpc. Even with this conservative limit, bars in galaxies with M$_\star <10^{10}$M$_\odot$  remain systematically longer than predicted by extrapolating the high-mass bar length–stellar mass relation to lower masses.
Separating galaxies according to the presence of a bulge component in their final decomposition models reveals distinct trends (Figure \ref{fig_relation_size_bar_mass_total}. The increase in absolute bar length with stellar mass above M$_\star =10^{10}$M$_\odot$ is driven primarily by galaxies with bulges, whereas the nearly flat relation below this mass is dominated by bulgeless galaxies. Thus, the change in the slope of the overall relation coincides with a change in the dominant galaxy population.
To account for differences in host galaxy size, we also examine bar lengths normalized by R$_{25.5}$\footnote{from \href{https://irsa.ipac.caltech.edu/data/SPITZER/S4G/overview.html}{https://irsa.ipac.caltech.edu/data/SPITZER/S4G}.} (Figure \ref{fig_relation_size_bar_mass_total}, bottom panel). Although the full sample shows no clear dependence of relative bar length on stellar mass, bulgeless galaxies exhibit increasing relative bar lengths toward lower masses. Bars in these systems therefore occupy a larger fraction of their host disks. This trend becomes less apparent when galaxies with bulges are included. 
In global terms, the relative bar length is nearly constant with stellar mass, which is similar to findings of other studies (see also \citealt{Wang25}).

Furthermore, when we restrict ourselves to the stellar-mass range over which the bulge and bulgeless subsamples overlap (9.5 $\leq \log($M$_\star/$M$_\odot)\leq$ 10.5), a second trend becomes apparent: at a given galaxy stellar mass, galaxies with a bulge component systematically exhibit larger relative bar sizes than their bulgeless counterparts. Thus, within this common mass range, the presence of a bulge is associated not only with a greater proportion of stellar-mass contained within the bar (Figure~\ref{fig_relation_mass_bar_model}, bottom panel), but also with relatively longer bars (Figure~\ref{fig_relation_size_bar_mass_total}, bottom panel). These results suggest a possible relation between these bar properties and the presence of central stellar structures, although this correlation does not by itself imply a causal connection between the presence of a bulge and bar growth.

\section{Discussion}
\label{Chapter5}

Our results point to trends between bar properties (i.e., relative size and stellar mass content) and galaxy stellar mass, where the presence or absence of a bulge component appears to bring about differences. We find that galaxies with no prominent bulges appear to display an interesting trend towards the low-mass end, with bars extending out further into their host disks (i.e., longer relative sizes for less massive galaxies; Figure~\ref{fig_relation_size_bar_mass_total}, bottom panel). We also find that higher-mass galaxies ($\sim$M$_\star > 10^{9.5} $M$_{\odot}$) with a bulge component in their final models (or predominantly nuclear disks, considering Sérsic indices of the majority of these bulge components) host bars with larger relative sizes and with a greater proportion of the galaxy stellar mass contained within them (Figures~\ref{fig_relation_mass_bar_model} and \ref{fig_relation_size_bar_mass_total}, bottom panels).

The results regarding bar size are consistent with studies that find longer bars in galaxies with more prominent central stellar structure \citep{Gadotti11,Hoyle11}, and with studies reporting longer bars in earlier-type spiral galaxies \citep{kmd+07, Erwin+18}.
The trends involving the stellar mass contained within the bar component is a new aspect that our work brings to light. Although a number of interrelated trends come into play (e.g., galaxy stellar masses, Hubble morphological type, bulge prominence), where fundamental relations become challenging to identify, our results both in size and stellar mass content within the bar structure suggest different bar evolutionary pathways for galaxies of distinct masses.

\citet{sheth+08} observationally showed that bars in more massive galaxies become established earlier than in their lower-mass counterparts, with lower-mass galaxies likely acquiring the conditions necessary to form and sustain a bar later on.
Within this scenario it appears to follow naturally that bars in more massive galaxies have had ample time (more so than lower mass galaxies) to grow in bar-to-total stellar mass ratio. The presence of bulges in the majority of barred disks at the higher-mass end may result from a combination of merger-driven bulge formation in early disks \citep{Hopkins10}, and secular evolution driven by the bar.
Bar-driven gas inflows toward the central regions \citep{Hawarden+1986,Ho1997} 
can build up a central gas reservoir (according to observational studies, \citealt{Sakamoto1999, Sheth+05}; and simulations, \citealt{Seo19,Verwilghen25}) contributing to the formation of a central stellar structure \citep{Bittner20,Gadotti2020,Schulthesis25} or to the further growth and transformation of a pre-existing bulge.

In low-mass galaxies, longer bars extend further into the disk and have larger relative sizes (Figure~\ref{fig_relation_size_bar_mass_total}, bottom panel), particularly in the absence of a prominent bulge. For some cases, the environment could play a role in triggering or enhancing bar formation. An example is found in \citet{Cuomo24}, where the authors studied the dynamical state -- including measurements of bar length -- of two dwarf galaxies in the Virgo Cluster: NGC 4483 ($7.5 \times 10^{9} M_\odot$; 2.2 kpc) and NGC 4516 ($5.7 \times 10^{9} M_\odot$; 2.9 kpc). They found that these galaxies host bars that are large relative to their galaxy size, when compared with the disk scale length. However, it is important to emphasize that this result was found for only two dwarf galaxies and should therefore not be extrapolated as a general interpretation of our results, particularly in the absence of environmental information for our sample. We mention it here only as a possible mechanism that could contribute to the formation of long bars in dwarf galaxies. Other studies investigating bar lengths across different environments have also reported contrasting results. Studies of early-type spirals \citep{Aguerri+23} and of high-mass galaxies \citep[$M_\star > 10^{10} M_\odot$;][]{Cuomo+26} found longer bars in less dense environments. These results highlight that our understanding of whether and how the environment affects bar size remains limited, while also suggesting that environmental effects may depend on galaxy stellar mass.
Together, these results indicate that bar formation and growth are influenced by multiple processes – including the epoch of formation, galaxy interactions, environmental density, and host galaxy mass – pointing to a complex evolutionary scenario in which bars can arise though different pathways.

\section{Summary and Conclusions}
\label{Chapter6}
We select a sample of well established bars in the nearby universe to quantify the stellar mass content within the bar structure and its relation with other galaxy global properties. For this, we use GALFITM to decompose 3.6 and 4.5 $\mu m$ images of 369 galaxies from the $S^4G$ sample. 
Our analysis provides a basis for measuring the stellar mass associated with each galaxy component.
The final decomposition models for our sample consist of two to four components. The simplest models include disk and bar components, while the most complex models incorporate a disk, bar, bulge, and a nuclear point source.
We successfully obtained final models for 344 out of the 369 galaxies in our sample. Our main results are summarized as follows:

\begin{itemize}

    \item 
    The majority (70\% of the 103 galaxies) of galaxies with a bulge component in their final decomposition model has a Sérsic index n $\leq 2$, consistent with the photometric signature of a nuclear disk
    \citep{Gadotti2009,Gadotti2011,Gadotti2020}. 
    This finding is consistent with studies that show a connection between bar-driven gas inflows toward central regions and the formation and growth of nuclear disks \citep{Seo19,Gadotti2020,Schulthesis25}.

    \item More massive galaxies ( M$_\star \geq  10^{10} $M$_\odot$) tend to host more massive bars, and galaxies with bulge component in their final decomposition models also tend to host massive bars. This trend is observed not only in absolute terms but also in relative terms, with bars in more massive galaxies with a central stellar structure encompassing $\sim 11\%$ of the galaxy stellar mass, compared to $\sim 7\%$ in their low-mass counterparts without bulge component.

    \item 
    Bulgeless galaxies appear to show a trend where bars extend out further into their host disks as these disks become less massive. When limited to the mass range over which the bulge and bulgeless subsamples overlap ($9.5\leq\log($M$_\star/$M$_\odot)\leq10.5$), we identify a bimodality in the relative bar size distribution, where at a given galaxy stellar mass, galaxies with a bulge component systematically exhibit larger relative bar sizes than their bulgeless counterparts.

\end{itemize}

Our results for the more massive galaxies could potentially point to a connection between bar growth and the presence of a central stellar component in more massive galaxies may act as an angular-momentum sink and thereby facilitate bar growth. Bar evolution in bar proprieties could be reflected both geometrically, through an increase in (relative) bar length, and dynamically, through an increase in the proportion of the galaxy's stellar mass becoming associated with the bar as stellar orbits are progressively trapped by the bar instability. However, our results on the high-mass end could also be a reflection of different disk formation epochs, where the presence of a bulge in massive galaxies merely points to the formation of a disk in a more distant past, with more time to grow the bar in size and stellar mass.
Although more detailed dynamical considerations need to be folded in to properly understand the underlying physics that gives rise to our observational results, this work suggests that different formation and evolutionary pathways may be at work, affecting the growth of bars in massive systems, while externally driven processes may become increasingly relevant toward lower masses.

\section*{Acknowledgements}
\begin{acknowledgments}
This work is based on observations made with the Spitzer Space Telescope, which is operated by the Jet Propulsion Laboratory, California Institute of Technology under a contract with NASA. Support for this work was provided by NASA through an award issued by JPL/Caltech. 
Y.C.C. acknowledges the support of the Coordenação de Aperfeiçoamento de Pessoal de Nível Superior - Brasil (CAPES) - Finance Code 001, the Brazilian National Research Council (CNPq grant 200135/2025-2) and the Serrapilheira Institute (grant Serra-1709-17357). K.M.D. thanks the support of the Serrapilheira Institute (grant Serra-1709-17357), the Brazilian National Research Council (CNPq grant 308584/2022-8) and the Rio de Janeiro Research Foundation (FAPERJ grant E-08/200.335/2026), Brasil.
T.K. acknowledges support from the Basic Science Research Program through the National Research Foundation of Korea (NRF) funded by the Ministry of Education (No. RS-2025-25399934).
A.C. acknowledges the grant 316868/2023-0 of CNPq.
K.S. acknowledges support by the French government under France 2030, Aix-Marseille University’s Initiative of Excellence-Amidex AMX-25-SPS-278.
M.G. acknowledges support from FAPERJ grant E-26/211.370/2021 and from CNPq grant 313720/2026-6.
\end{acknowledgments}

\software{GALFITM \citep{Bamford2011,Vika+14,Haussler+22} software based on GALFIT \citep{Peng+02,Peng+10}, Python \citep{python}, Astropy \citep{astropy13,astropy18,astropy22}, 
Matplotlib \citep{matplotlib07}, Scipy \citep{mckinney-proc-scipy-2010}, Pandas \citep{pandas_21500199}, Jupyter Notebook \citep{kluyver2016jupyter}
}

\bibliography{sample631}{}
\bibliographystyle{aasjournal}

\appendix

\section{The Need for 4.5~\micron~  for Component Stellar Mass Estimates}

\label{Appendix_Underestimating}

Our work aims to estimate the stellar mass distribution among the different morphological components of barred galaxies based on the 3.6 and 4.5~$\mu$m bands. The decomposition in the 3.6~$\mu$m band is provided by the $S^4G$ collaboration and is publicly available in \citet{Salo+15}.
We could, in principle, considering only the 3.6~$\mu$m decomposition, estimate the 4.5~$\mu$m flux distribution among the different components assuming that the relative flux contributions of the components remain the same in both bands (i.e., if the disk component accounts for 80\% of the total model flux at 3.6~$\mu$m, we assume that it also accounts for 80\% of the model flux at 4.5~$\mu$m).  This approach essentially freezes the components' individual [3.6 - 4.5] colors to follow the global galaxy color. Here, we investigate the impact of this assumption on the stellar mass estimates of individual components. Using the total galaxy magnitudes reported by \citep{Munoz_Mateos+15}, we can estimate the flux of each component at 4.5~$\mu$m even when only the 3.6~$\mu$m decomposition is available.

\begin{figure}[ht]
  \centering
  {\footnotesize
   \includegraphics[width=0.45\textwidth]{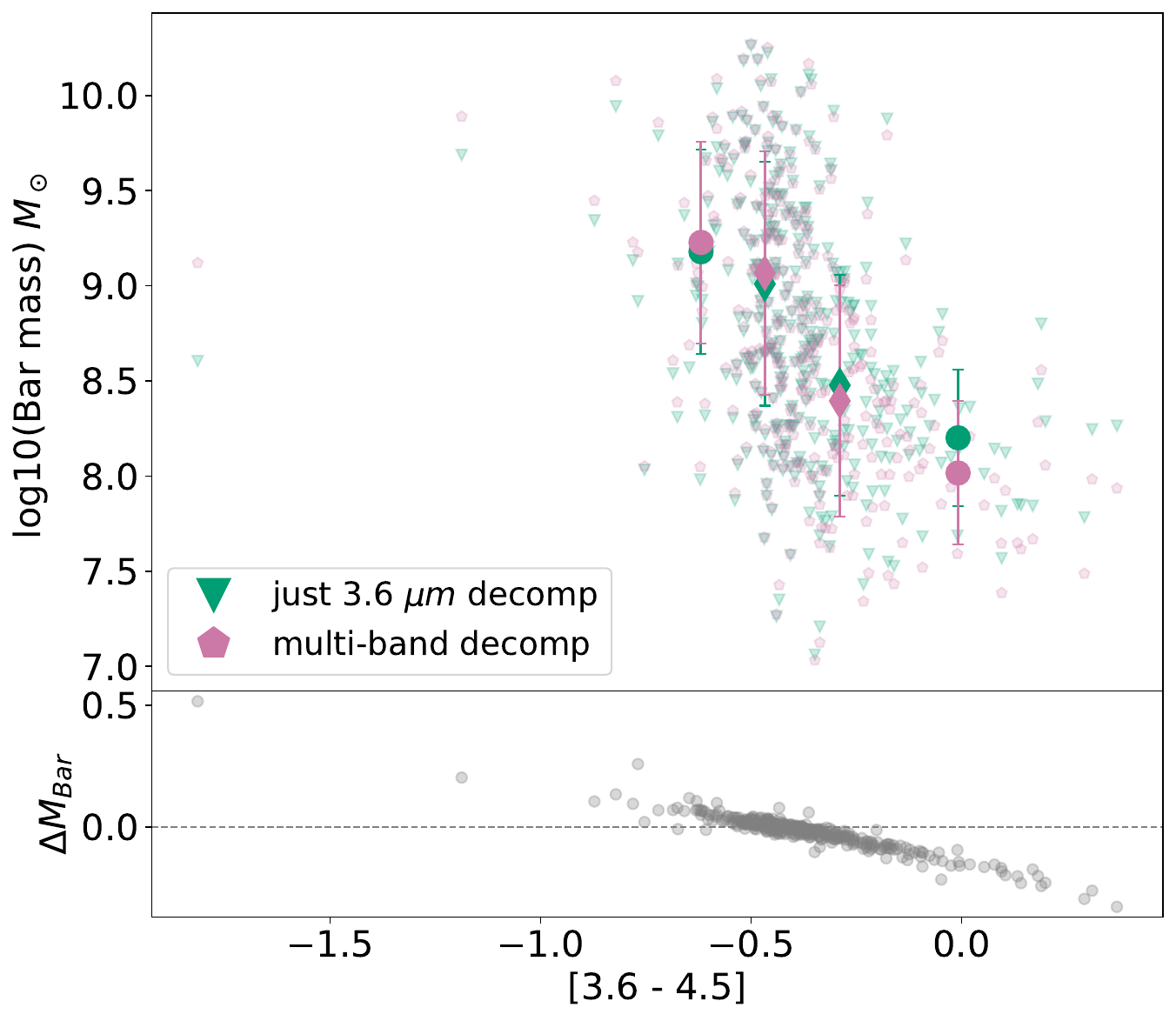}
    \caption{The main panel (above) shows the bar color (considering the bar fit in both bands) and the bar mass; the below panel shows the bar mass residual (bar mass calculated using the multi-band decomposition minus the bar mass calculated considering the decomposition in both bands equivalent). The green triangle indicates the bar mass calculate with just the decomposition in 3.6 $\mu m$, while the pink pentagon indicates the bar mass calculated with the multi-band decomposition. For the bigger symbols, the diamonds are the medians when the sample is divided in half, while circles are the bars 10\% more reddest or bluest (representing 34-35 galaxies from the edge region in this figure).}
    \label{fig_BARS_color_residual}   
    }
  \end{figure}

Based on \citet{Querejeta+15} we calculate stellar masses for each component using on the one hand the 3.6 and 4.5 $\mu m$ fluxes resulting from the explicit decomposition of images in both bands (which is the multi-band decomposition this paper proposes) and on the other hand using the 3.6 and 4.5 $\mu m$ resulting from the sole decomposition of the 3.6 $\mu m$ band.
Considering that the explicit decomposition of images in both the 3.6 and 4.5 um bands allows for each component to have its own independent color (and that restricting the decomposition to the 3.6 $\mu m$ band fixes a global color for all components), we probe for any trends between component stellar mass and component color. Figure \ref{fig_BARS_color_residual} shows the stellar mass of the bar component as a function of color, where the stellar mass is derived from the 3.6 and 4.5 $\mu m$ fluxes resulting from the multi-band decomposition (shown as pink pentagons), as well as the one derived using only the 3.6 $\mu m$ decomposition (down-pointing green triangles).
In the lower panel of Figure~\ref{fig_BARS_color_residual}, $\Delta M_{bar}$ shows the difference in the bar mass calculated using the multi-band decomposition and the 3.6 $\mu m$-only decomposition. This shows that bars with the reddest [3.6 - 4.5] colors tend to have lower stellar masses with the multi-band decomposition approach than with the single-band decomposition. Meanwhile, the stellar mass in bluer bars tends to be higher using the multi-band decomposition. 

\begin{figure}[ht]
  \centering
  {\footnotesize
   \includegraphics[width=0.45\textwidth]{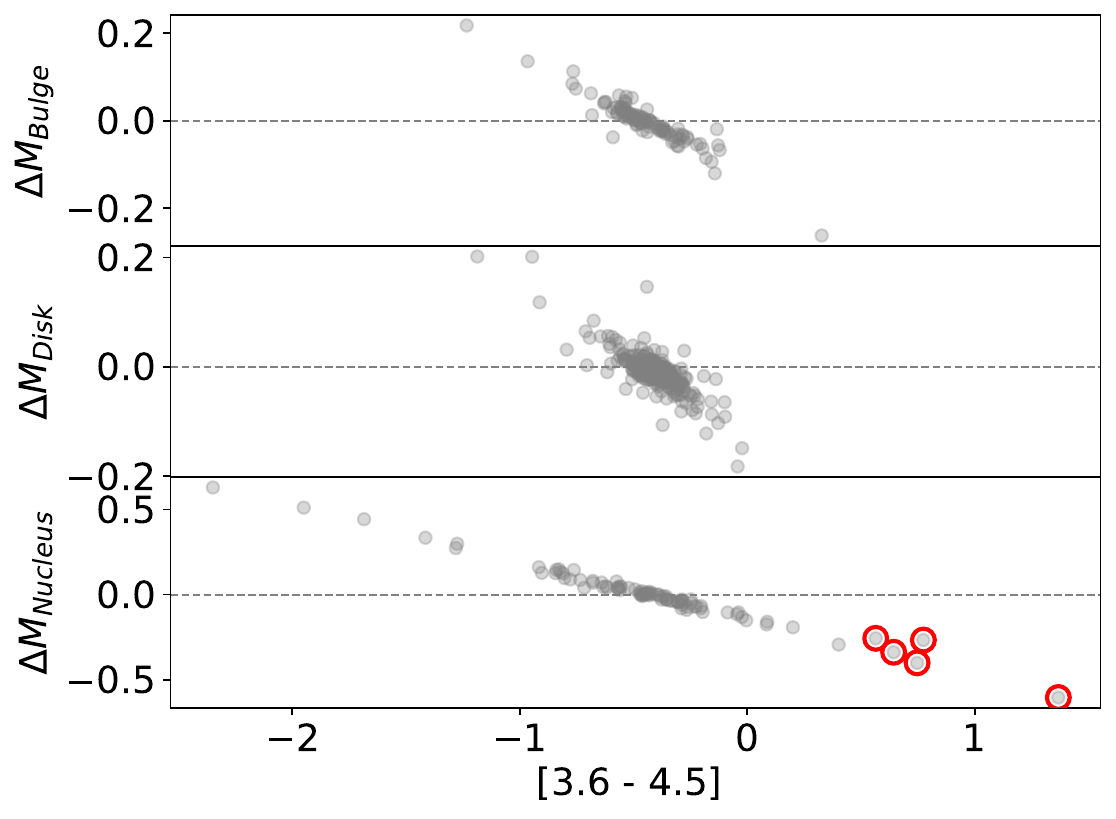}
    \caption{The residual stellar mass (i.e., component stellar mass using the multi-band decomposition minus the component mass calculated just with 3.6 $\mu m$ decomposition) to bulge (above), disk (middle) and nucleus (below). For the nucleus panel, the red circles indicates AGNs identified in \citep{Ikiz+20}.}
    \label{fig_3panels_color_AGN}    
    }
  \end{figure}

This suggests that relying solely on the 3.6~$\mu m$ band decomposition, which forces a galaxy-wide color onto each individual component, results in the stellar mass of the redder bars to be overestimated and that of the bluest bars to be underestimated. We find a similar trend when inspecting the remaining components (Figure~\ref{fig_3panels_color_AGN}: bulge, disk, nucleus). A particularly revealing case is that of nuclear point sources. We identify at the reddest end, five galaxies that have been identified as hosting a central AGN (NGC~7418; NGC~5347; NGC~4293; NGC~4593; NGC~1365; \citealt{Ikiz+20}). The presence of an AGN is expected to enhance the underlying 4.5 $\mu m$ dust continuum \citep{Galliano2018}, while a central starburst region is expected to enhance the 3.6~$\mu m$ detected flux due to 3.3~$\mu m$ PAH band contamination of the 3.6~$\mu m$ band. A range in colors is thus naturally expected for these nuclear point source components. This exemplifies quite clearly the need for the inclusion of the 4.5~$\mu m$ decomposition to guarantee improved stellar mass estimates for individual components.

\section{Preparing ingredients for the decomposition: sky, noise and PSF characterization}
\label{preparing ingredients}

To perform the fit, GALFITM requires a set of four input images for each galaxy: the science image, a mask image to exclude foreground stars and background galaxies, a sigma image to represent the error pixel values, and the point spread function (PSF) for each band used.  
The science and mask images are provided by the $S^4G$ collaboration as part of the Pipeline 1 and Pipeline 2 products, respectively. 
The sigma image represents a noise characterization for the science image, and to create it we follow the methodology described by \citet{Salo+15}. 
As an initial step in preparing the images for decomposition, we subtract the sky background level from the science image and calculate the sky root mean square (RMS), which is required for the construction of the sigma image.
We estimate the sky background level and the sky RMS by analyzing an arc-shaped region located away from the galaxy (i.e., between 6.3 and 8.1$^{\prime}$ from the galaxy center). The galaxy center is determined using the information provided in the image header. 
We mitigate the extreme pixel count variations associated to the image edges, by considering only the central 90\% of the pixel values within this sky region (i.e., from the 5th to the 95th percentile), excluding extreme values when computing both the sky background level and its RMS. 
All these calculated values are provided in tabular form in the GitHub repository associated with this work\footnote{\href{https://github.com/Cavalcante-Coelho/Stellar-Mass-Census-of-Local-Bars}{https://github.com/Cavalcante-Coelho+26.}}.

The sigma image provides an error value for each pixel, which is necessary to fit analytical functions and to create a decomposition model for the galaxy image.
\citet{Salo+15} provided sigma images for the 3.6$~\mu m$ band, however, a sigma image for the 4.5~$\mu m$ band is also required in our analysis. To ensure uniformity in our approach, we construct the sigma images from scratch for both the 3.6 and 4.5~$\mu m$ bands.
The sigma image has two primary contributions: the noise associated with the photon count arriving at the instrument and the noise inherent to the instrument itself. 
Photon noise follows a Poisson distribution and is associated with either the galaxy light or the sky background.
In the science image, the pixel values are expressed in ADU (Analog-to-Digital Unit) and are related to the number of electrons in each pixel. For the analysis, it is important to determine how many electrons correspond to a given pixel value. This information is typically provided by the ``Gain" keyword in the image header. 
The initial step in creating the sigma image is to recover the number of electrons, $N_{e}$, associated with each pixel. For this, we use

\begin{equation}
    N_{e} = \frac{F + F_{bg}}{F_{conv}} \times T_{frame} \times N_{frames} \times g.
    \label{eq_Ne}
\end{equation}

\noindent
where $F$ represents the pixel values in the science image, $F_{bg}$ is the zodiacal light (header keyword SKYDRKZB), $F_{conv}$ is the conversion factor between flux units and the original digital units (header keyword FLUXCONV, units MJy sr$^{-1}$ per DN~s$^{-1}$, where DN denotes the digital number recorded by the detector), $T_{frame}$ is the integration time per frame, measured in seconds, $N_{frames}$ is the number of combined frames contributing to each pixel value, and $g$ is the detector gain factor (header keyword GAIN, units electrons per DN). We calculate the number of combined frames for each pixel following the methodology described by \citet{Salo+15}, using $N_{frames} = W/10$, where $W$ represents the pixel values of the weight image provided by the publicly available reduced images in the 3.6 and 4.5 $\mu m$ bands \citep{Sheth+10}, which are given in units of MJy~sr$^{-1}$.

The next step is to calculate the statistical uncertainty in the number of electrons. This uncertainty is computed by considering both the photon noise and the detector readout noise (RON).

\begin{equation}
    \sigma^{2}(N_{e}) = N_{e} + N_{franes} \times RON^{2}
    \label{eq_sigma_Ne}
\end{equation}

We adopt RON values of 15.0, 14.6, and 21.0 electrons for header keyword FRAMTIME values of 12, 30, and 100 s, respectively. These values were provided by the Spitzer Science Center Helpdesk and differ slightly from those listed in the image header keyword RONOISE \citep{Salo+15}. Therefore, we use $\sigma(N_e)$ to estimate the flux uncertainty of the image

\begin{equation}
    \sigma_{est}(F) = \sigma{N_{e}} \times \frac{F_{conv}}{T_{frame} \times N_{frames} \times g}
\end{equation}

The $S^4G$ survey was observed during the Spitzer warm mission, but some images were incorporated from objects that had already been observed during the cryogenic mission. In \citet{Salo+15}, the relationship between the RMS of the sky in each image and the value of $\sigma_{est}$ was investigated. Reasonable values for the sigma image are those close to the sky RMS of each image. For the cryogenic mission phase, the RMS is approximately $0.9\sigma_{est}$

\begin{equation}
    \sigma_{image}(F) = 0.9 \sigma_{est}(F) \;\;\;\;\;\;\;\;\;\;\;\; \text{(cryogenic mission)}.
\end{equation}

However, for images from the warm phase, the observed RMS is approximately 50\% higher than the theoretically estimated value, indicating the presence of additional noise. The Spitzer Science Center Helpdesk suggested the inclusion of an instrumental noise component ($\sigma_{conf}$), which is added quadrature to the theoretical noise estimate. To estimate this additional noise, we consider

\begin{equation}
    \sigma_{conf} = \sqrt{\left(\frac{RMS_{sky}}{0.9}\right)^{2} - median(\sigma_{est})^{2}}
\end{equation}

\noindent
Finally, we estimate the value for the warm mission

\begin{equation}
    \sigma_{image}(F) = 0.9 \sqrt{ \sigma^2_{est}(F) + \sigma^2_{conf}}  \;\;\;\;\;\;\;\;\;\;\;\; \text{(warm mission).}
\end{equation}

Although the $S^4G$ decomposition of 3.6 $\mu$m band presented by \citet{Salo+15} only used $\sigma_{conf}$ for galaxies from the warm mission, we calculate this value for all galaxies in our sample, with some values being zero. Given the observed relationship between the sky RMS and the sigma image value, we choose to apply a $\sigma_{conf}$ correction when necessary (i.e., when $\sigma_{conf}$ $>$ 0).

A representative PSF is crucial for a consistent photometric decomposition. Central components, particularly the bulge and the nuclear point source, are highly sensitive to the adopted PSF. Parameters such as the bulge effective radius and Sérsic index, as well as the magnitude of the nuclear point source, can vary significantly depending on the PSF used in the fit. 
To create our PSF, we select stars from the galaxy images in our sample using the \texttt{find\_peaks} function from the \texttt{Photutils} library in Python.
The source selection consists of:
(1) an automated selection: isolated sources with peak values higher than 30 $MJy/sr$ to ensure sufficient signal to noise, and with a distance from the galaxy over more than 100 pixels (1.25$^{\prime}$); and (2) a visual inspection, in which elongated, distorted,
or saturated sources are excluded, selecting only the best sources.
With the sources selected, the \texttt{EPSFBuilder} from the \texttt{Photutils} library is used to generate the PSF, adopting an oversampling factor of 4. 
This is necessary because images from the $S^4G$ survey have a limited spatial resolution (i.e., the pixels do not ideally capture fine details of the image), with the resolution of the IRAC camera being about 1.2". This value is close to the dispersion of a point source observed in these channels, making this a case in which the use of an oversampled PSF is recommended \citep{Peng+10}.
The number of best selected sources used to build the PSFs for the 3.6 and 4.5 $\mu m$ band are 25 and 27 stars, respectively, with corresponding FWHM values of 1.89$^{\prime\prime}$ and 1.79$^{\prime\prime}$.

Considering the initial guess parameters based on previous work from \citet{Salo+15} and all the ingredients prepared, we proceed to the final steps required to perform the decomposition.  
The IRAC instrument, by construction, was set up to provide two fields of view at each band: one with the target and one with an adjacent sky view. The targets are placed in opposite positions in 3.6 and 4.5 $\mu m$ (see IRAC Instrument Handbook for details). 
As a result, the observed galaxy is not centered on the central pixel of the image, as well as the pixel position of the galaxy center varies significantly between bands. In addition, the images have pixels with Not-a-Number (NaN) values or values that do not represent observed data, mostly at the edge of the image. 
For this reason, we crop the images to a reasonable size that encompasses the galaxy and center the galaxies in both bands in the same pixel value, an important step using GALFITM. These images cover three times the semi-major axis measured at 3.6 $\mu m$ (25.5 AB mag/arcsec$^2$), as obtained by \citet{Munoz_Mateos+15}. 
Finally, we subtract the sky background from the cropped science images.

\section{Recovery of galaxy stellar masses with GALFITM}
\label{Comparison galaxy and model stellar mass}

\begin{figure}[ht]
  \centering
  {\footnotesize
   \includegraphics[width=0.5\textwidth]{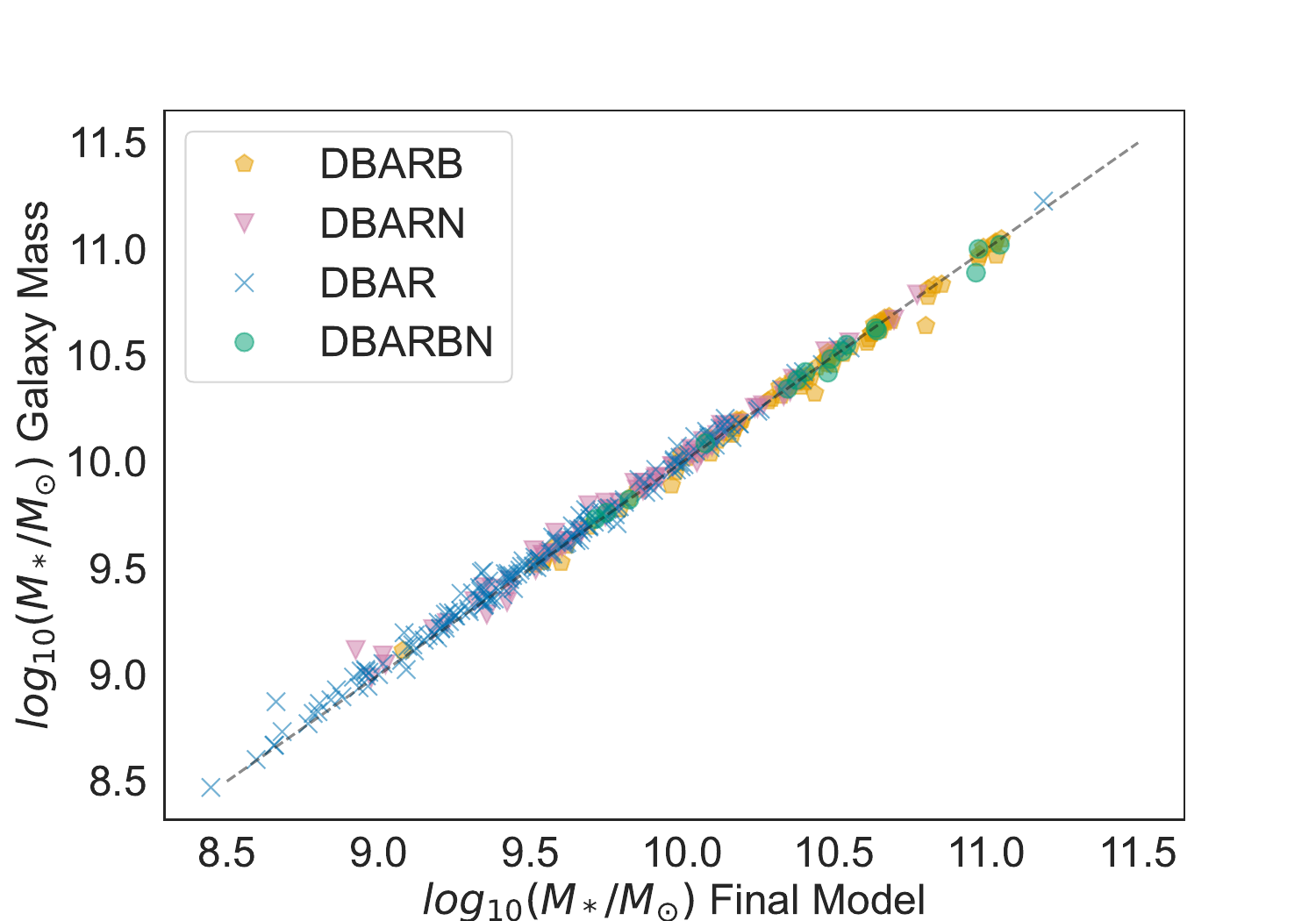}
    \caption{Comparison of the total stellar mass derived from our final decomposition models with the galaxy stellar mass calculated using the magnitudes from \citet{Munoz_Mateos+15}. The figure highlights, in the legend, the different final decomposition models considered in our analysis.}
    \label{fig_mass_comparison}   
    }
  \end{figure}

We calibrate the fluxes of our components using Equations \ref{eq_Fdecomp} and \ref{Fcomp} for both the 3.6 and 4.5 $\mu$m bands. This correction accounts for the residual flux while preventing the inclusion of excess model flux, which can arise from GALFITM smoothly extending the fitted light distribution into the outer regions. We then calculate the stellar mass of each component using these corrected fluxes. The final model mass is defined as the sum of the masses of all components included in each galaxy model.
Figure \ref{fig_mass_comparison} shows our sanity check for the stellar mass in our final models, compared to the galaxy stellar mass (i.e., galaxy mass calculate using \citealt{Querejeta+15} considering the magnitude estimates from \citealt{Munoz_Mateos+15}, where the values are measured within concentric ellipses down to a surface brightness level of 25.5 AB $mag$ $arcsec^2$).  
In general, we find good agreement between the mass estimates, where the differences are mainly due to the residual flux. In some cases, we find negative residuals, typically associated with complex structures for which GALFITM has difficulty converging to the best model; however, overall, this does not affect our final results. 

\section{Robustness of bar mass estimates}

\label{appendix_barwithresidual}
\begin{figure*}[ht]
  \centering
  {\footnotesize
   \includegraphics[width=0.8\textwidth]{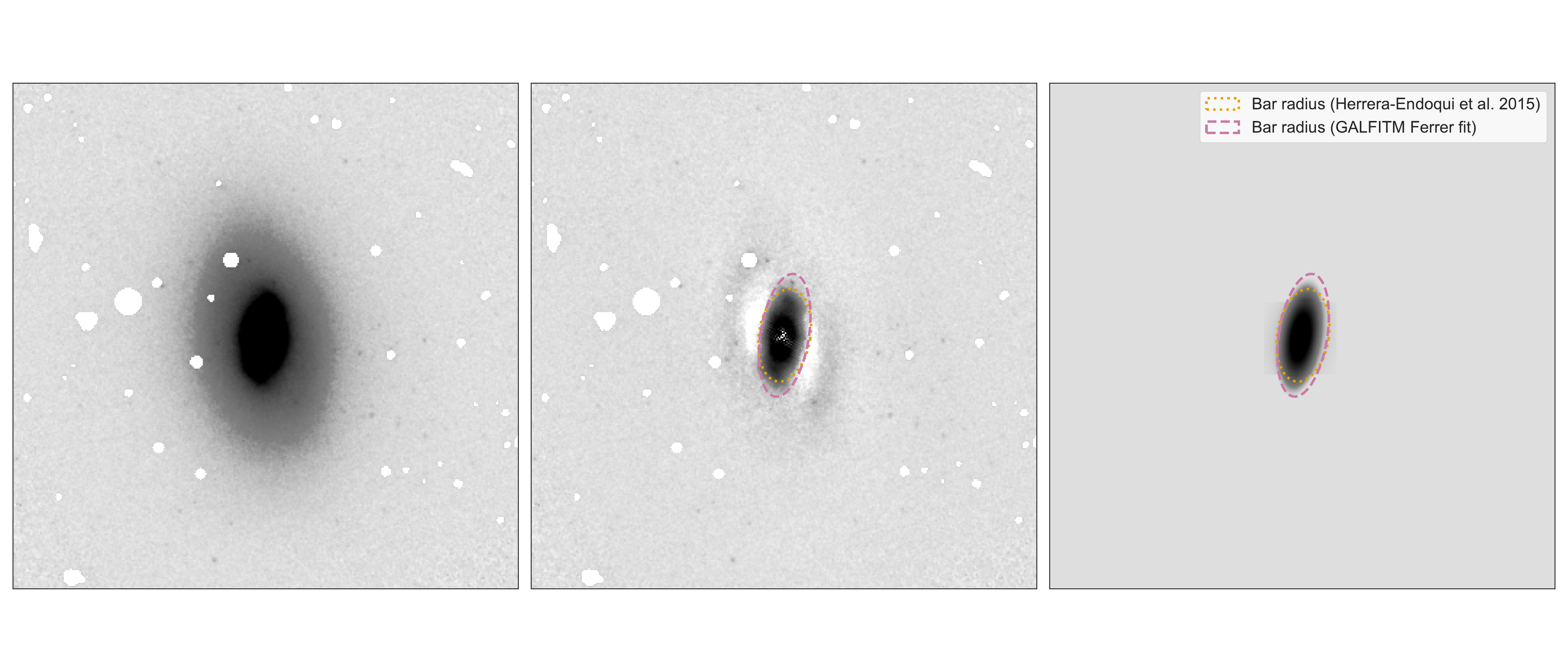}
    \caption{NGC 3941 in the 3.6 $\mu$m band science image with mask for foreground stars (left); the residual plus bar light image (middle) as an example of the bar light region used to calculate the bar stellar mass; and bar model image (right). The yellow ellipse is the bar length from \citet{herrera-endoqui2015} and in pink is the bar length from the decomposition output (i.e., bar truncation radius). 
    }
    \label{bar_area_residual}    
    }
  \end{figure*}

The Ferrer function is well suited for representing bars in massive galaxies, where a flat bar surface brightness profile is expected \citep{Kim15,Erwin2023}. 
We calculate the stellar mass in bars considering three approaches: (i) the light contained within the Ferrer bar component resulting from the fit; (ii) the light within the bar region, as defined by the Ferrer truncation radius, measured in the galaxy image that has been masked for foreground stars and where all non-bar components of the final model   (i.e., disk, bulge and/or nucleus) have been removed;
 and (iii) the light contained within the bar region in the same residual image as described in (ii,) but where the bar region is defined by the visual bar lengths provided by \citet{herrera-endoqui2015}.
We note that the bar sizes derived from the Ferrer function fits (the bar truncation radius) tend to be longer than the values obtained from the visual measurement by \citet{herrera-endoqui2015}. 
The Ferrer model is designed to reproduce the bar edges smoothly, with the surface brightness gradually decreasing to negligible values and reaching zero at the outer boundary of the component. Consequently, the bar length inferred from our fit tends to be longer than that obtained from ellipse fitting or visual inspection.
Figure \ref{bar_area_residual} shows an example of the bar region considering the bar length of the fit and the visual measurement of the bar length from \citet{herrera-endoqui2015}, in a bar residual image (i.e., galaxy image minus other components) and bar model image.

Although the bar lengths derived from our fits are generally longer than the visual estimates of \citet{herrera-endoqui2015}, adopting the smaller visual bar lengths to calculate the bar masses does not change the observed trend. This is shown in Figure \ref{fig_relation_mass_bar_model}. The red line shows the fit obtained using the bar mass derived directly from the Ferrer-model bar light, while the black dashed line shows the fit obtained when the bar mass is calculated from the residual image (galaxy image that has been masked for foreground stars and where all non-bar components of the final model have been removed) within the bar extent defined by \citet{herrera-endoqui2015}. Both line fits point to the same trend.

\begin{figure}[ht]
  \centering
  {\footnotesize
   \includegraphics[width=0.5\textwidth]{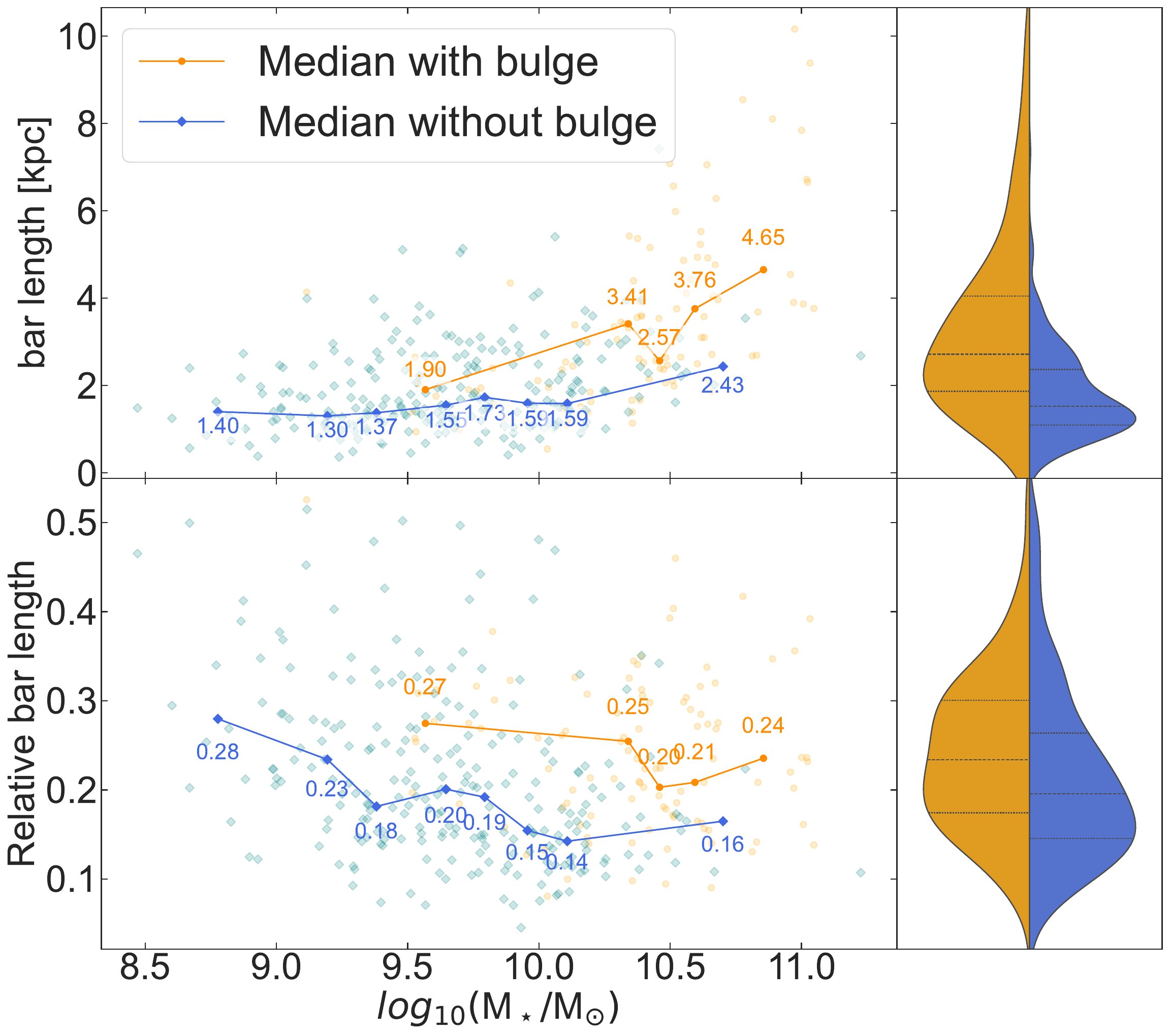}
    \caption{The relation between the bar size from \citet{herrera-endoqui2015} and the relative bar size, and the galaxy stellar mass for our sample considering galaxies with bulge (yellow) and without bulge (blue) in its final model.}
    \label{herrera_barsize_with_without_bulge}   
    }
  \end{figure}

We also tested the relation between galaxy mass and bar length as defined by the Ferrer truncation radius (shown in Figure \ref{fig_relation_size_bar_mass_total}), but using the visual bar lengths from \citet{herrera-endoqui2015}. This is shown in Figure \ref{herrera_barsize_with_without_bulge}, where the same overall trend as Figure \ref{fig_relation_size_bar_mass_total} is recovered, where bar length increases toward higher galaxy masses, particularly for galaxies that contain a bulge component.

\end{document}